\documentclass[a4paper,11pt]{article}
\pdfoutput=1 

\usepackage{jcappub} 
\usepackage{graphicx}
\usepackage{multirow} 
\usepackage{amssymb}
\usepackage{amsmath}
\usepackage{enumerate}
\usepackage{subcaption} 
\usepackage{color} 
\usepackage[normalem]{ulem}
\usepackage[T1]{fontenc}
\usepackage[utf8]{inputenc}
\usepackage[american]{babel}

\newcommand{\xmax}{\ensuremath{X_{\rm max}\,}}

\newcommand{\zhaires}{\mbox{ZHA{\scriptsize{${\textrm{IRE}}$}}{S }}}

\title{Revisiting the Radio Lateral Distribution Function: An amplitude dependence on \xmax and primary composition}

\author[1]{Washington R. Carvalho Jr.\note{Corresponding author.}}
\author{and Lech Wiktor Piotrowski}
\affiliation{Faculty of Physics, University of Warsaw, Ludwika Pasteura 5, 02-093 Warsaw, Poland}
\emailAdd{carvajr@gmail.com}

\abstract{

  In this work we show that there is a strong dependence of the radio lateral distribution function (LDF) electric field amplitudes at ground level on the position of the shower maximum (\xmax) in the atmosphere, even after accounting for differences in the electromagnetic (EM) energy of the showers. Since an \xmax dependence also leads to a primary composition dependence, this implies that information on the primary mass composition is encoded not only in the LDF shape but also in its amplitude.  This \xmax dependence can be explained in terms of two competing scalings of the measured electric field: One goes with $(1/\rho)^J$, where $\rho$ is the air density at \xmax and $J$ is a zenith dependent non-linearity factor describing coherence loss. This density scaling tends to decrease the geomagnetic emission of deeper showers. The other scaling goes with $(1/R)$, where $R$ is the distance from \xmax to the core at ground, and instead increases the measured electric field of deeper showers. At low zenith angles, the $(1/R)$ scaling is stronger and leads to larger measured electric fields as \xmax increases. The picture at higher zenith angles, i.e., lower densities, is more nuanced. In this region, the deflections due to the Lorentz force are much larger and introduce extra time delays between the particle tracks, decreasing the coherence of the emission. This loss of coherence is highly dependent on the strength of the geomagnetic field and can slow down, or even reverse the increase of the radio emission with decreasing air density. This strong, yet historically overlooked additional dependence of the LDF amplitude on \xmax/composition could be used to directly infer, even bypassing any \xmax reconstruction, the cosmic ray primary composition on an event-by-event basis. It could also have some repercussions on other radio reconstruction methods, such as a possible \xmax/composition dependence in the estimation of the EM shower energy.}

\begin{document}
\maketitle
\flushbottom
\section{Introduction}

When a cosmic-ray primary particle enters the atmosphere, it interacts with an air nucleus, initiating a cascade of secondary particles known as an Extensive Air Shower (EAS). As the shower develops, a large number of relativistic secondary particles are produced. These particles propagate towards the ground at nearly the speed of light, forming a thin and dense disk called the shower front. EAS can be detected using a variety of techniques: surface detectors (SD) arrays that measure the secondary particles reaching the ground; fluorescence telescopes (FD), which observe the ultraviolet fluorescence emission created by the excitation of atmospheric nitrogen during the shower development; and radio detection (RD), which uses antenna arrays to measure the coherent electromagnetic emission generated by the shower as it develops. This radio emission is primarily due to the deflection of charged particles in the geomagnetic field (geomagnetic emission~\cite{kahnlerchegeo}), with a smaller contribution from an excess of electrons in the shower, produced as the shower front entrains electrons from the surrounding medium (Askaryan emission~\cite{Askaryan62}). All these detection techniques provide access to EAS observables. From these, one can infer the properties of the primary cosmic-ray particle, such as its energy, arrival direction, and mass composition, the latter being arguably the most challenging to determine. Due to the physics of shower development, the mass of the primary particle strongly influences the atmospheric depth at which the shower reaches its maximum (\xmax). Consequently, most reconstruction methods use \xmax as a surrogate for composition. In fact, all radio and fluorescence mass reconstructions rely, directly or indirectly, on \xmax. 

The radio detection technique was initially proposed in the 1960’s, but its development stalled in the mid 1970's due to technological issues. With the advent of modern digital signal processing, the development of the radio technique started again in the early 2000's~\cite{TimRadioRenaissance}. It is now a mature, well-established technique used by several experiments worldwide. In this work we investigate the dependence of the electric field amplitude at ground level on the position of \xmax in the atmosphere. It is well known that several factors influence the radio emission of extensive air showers, but a physically coherent explanation combining them to describe the dependence of the peak electric field amplitude on \xmax has not yet been presented. In the 1960s, the concept of the radio emission being linked to a transverse current due to geomagnetic deflection~\cite{kahnlerchegeo} was already established, giving rise to the well-known modern $|\vec{E}|\propto B\sin{\alpha}$ dependence, where $\alpha$ is the angle between the shower axis and the geomagnetic field $B$. A few years later, ``Allan's formula'' (Eq. 84 in \cite{Allan71}) formalized the dependence of the emission on primary energy and the distance to the shower, the latter implicit in the $\cos{\theta}$ term that accounts for the shower geometry, where $\theta$ is the zenith angle. Later, it was recognized that the electromagnetic (EM) component of the shower is, by far, the main contributor to the emission \cite{FalckeGorham,zhaires-air}. This made clear that the relevant energy scale for the radio emission is in fact the EM energy of the shower, instead of the total primary energy. More recently, it was also established that the atmospheric density at \xmax is a relevant factor affecting the strength of the geomagnetic emission \cite{scholten-driftvelocity}. Recent works have, to some extent, incorporated some of these factors into $X_{\rm max}$-based correction factors to improve the accuracy of energy reconstructions \cite{FelixTimInclinedRec,LukasInclinedICRC2025}. However, these approaches rely on multiple separate simulation-based parametrizations, without providing a full physical interpretation of how the atmospheric density at, and the distance to \xmax affect the peak E-field amplitudes. Recently, several papers have also addressed a loss of coherence effect that occurs at lower air densities and/or in the presence of higher geomagnetic fields~\cite{JuanLossCoherence,ChicheLossCoherence,GuelfandLossCoherence}. In this work we use the ZHS formalism~\cite{ZHS92,TimeDomainZHS} as a guide to the underlying physics and combine all these elements, including loss of coherence effects, to provide a complete and physically based picture of how the position of \xmax in the atmosphere influences the electric field amplitudes at ground level.

This paper is organized as follows: On section \ref{sec:motivation} we describe the motivation for this work, which was based on an analysis of the results of a simple Random Forest event-by-event mass discrimination method we have developed. On section~\ref{sec:zhairessims} we describe the radio emission simulations used in this work and on section \ref{sec:radioemission} we review the radio emission of air showers and also discuss how the position of the shower maximum in the atmosphere could influence the electric field amplitudes. On section~\ref{sec:Rrhoscaling} we describe the distance and air density scalings we used to explain and quantify the electric field amplitude dependence on \xmax, as well as the loss of coherence effect. On section~\ref{sec:PredictionsAndComparison} we make predictions for the behavior of the electric field amplitudes, based on shower geometry and the two previously discussed scalings. We then compare these predictions with the results of the full simulations. Finally, on section~\ref{sec:discussion}, we discuss some possible impacts this amplitude \xmax dependence could have on the radio detection field and present the conclusions.

\section{Motivation}
\label{sec:motivation}

The motivation for this work arose during the initial stages of development of a mass discrimination method for cosmic ray events using the radio detection technique~\cite{RFDiscrimination-ICRC2025,MLpaper}. This method uses supervised machine learning (ML) algorithms, namely random forests, to discriminate between light (p) and heavy (Fe) primary compositions, on an event-by-event basis. It bypasses any \xmax reconstructions and instead tries to infer the primary composition directly. To train and test the accuracy of this ML algorithm, we used radio events generated by RDSim~\cite{RDSimICRC,RDSimARENA,RDSimECRS} on a ground antenna array. For each zenith angle and energy, we generated thousands of proton- and iron-induced events with random core positions and azimuth angles. The generated events were divided into 2 sets, one for training the random forest, and the other for testing its accuracy. At this early stage, as features of the random forest we only used, for each triggered antenna, its distance to the shower axis and the peak amplitude of the electric field. Also, in order to account for energy measurement uncertainties and the intrinsic differences in the electromagnetic (EM) energy of showers induced by different primary compositions, we initially used a huge 30\% energy smearing for every event\footnote{Later in the development, we scaled the amplitudes of all simulations by the EM energy of each shower. So, in essence, all of our events would have the same EM energy instead of primary energy, removing the EM energy dependence on the mass of the primary particle. That allowed us to lower the energy smearing to 10\%, which is still more than twice the expected EM energy uncertainty of current radio EM energy reconstructions. See section \ref{sec:zhairessims} for more details.}. Even with such energy uncertainties, we were able to obtain accuracies of up to 82\% on the primary composition discrimination.

\begin{figure}[htb]
  \begin{center}
    \includegraphics[width=0.48\textwidth]{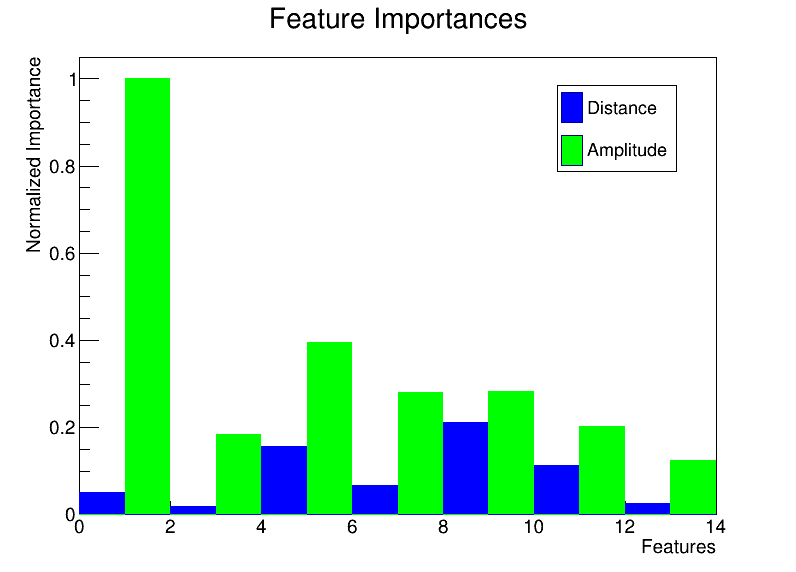}\includegraphics[width=0.52\textwidth]{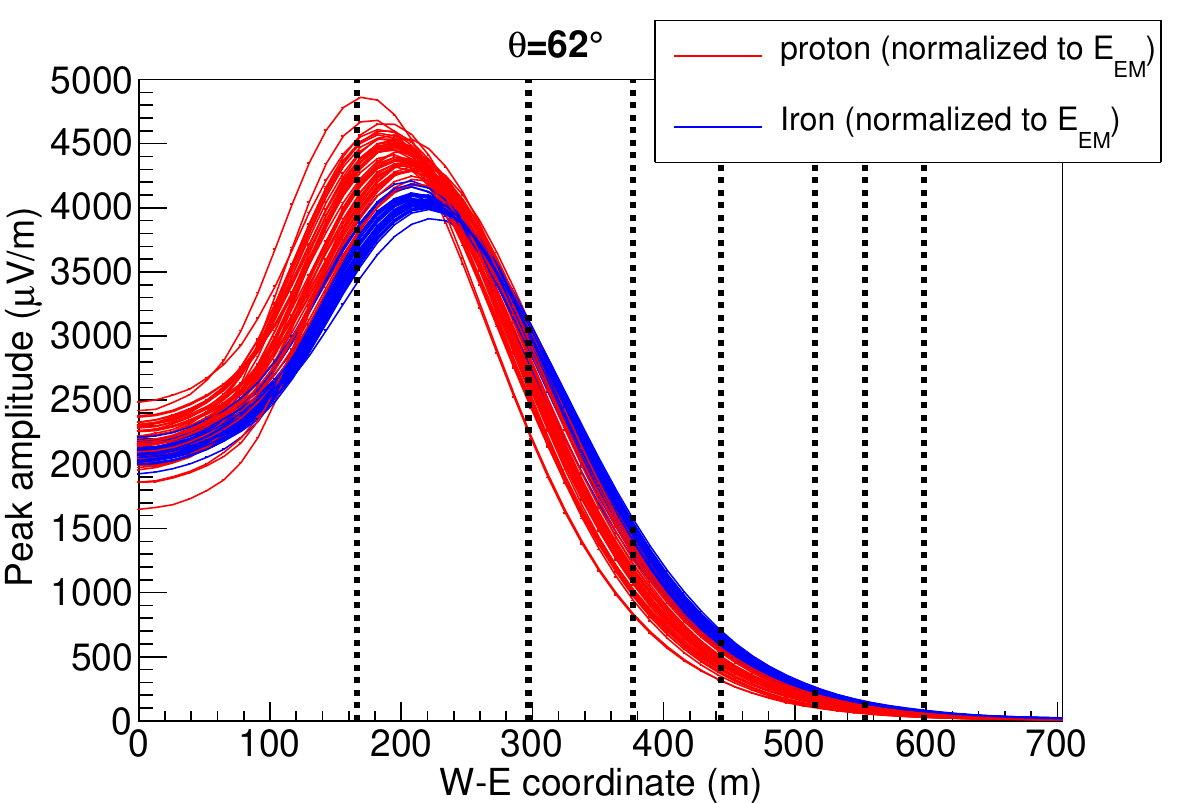}
    \caption{Left: Random forest normalized feature importances for events with zenith angle $\theta=62^\circ$ at the Giant Radio Array for Neutrino Detection (GRAND) site. For each triggered antenna in an event, there are two features: the distance to the axis (blue) and the peak amplitude (green). The antennas are ordered in increasing distance to the axis (from left to right). Right: LDFs obtained from full ZHAireS simulations of events coming from the North with $\theta=62^\circ$ at GRAND, normalized by the EM energy of each shower. These simulations were the ones used as input for RDSim in order to generate the training and testing events for the random forest. Note that RDSim morphs these input simulations to generate events with different azimuth angles and core positions. The vertical dashed lines, starting from the left, represent the average distance, for all events, to the closest antenna (first line), second closest antenna (second line) and so on. See text for more details. }
    \label{fig:featureimportances}
  \end{center}
\end{figure}

In order to understand how such a simple random forest could yield such good accuracies using just the antenna distances and field amplitudes, we performed an analysis of the forest's feature importances. This analysis hinted that, for most geometries, proton showers tended to have much higher fields on the inner part of the Cherenkov cone, closer to the core, if compared to iron-induced showers. An example of such analysis is shown on Fig.~\ref{fig:featureimportances}, for a set of events with zenith angle $\theta=62^\circ$. On the left panel we show the normalized feature importances, ordered in increasing antenna distance to the shower axis. The most important feature in this example was the amplitude of the closest antenna, followed by the amplitude of the third closest antenna and diminishing for increasingly more distant antennas.

On the right panel of Fig.~\ref{fig:featureimportances} we show the LDFs obtained from the 100 full ZHAireS simulations~\cite{zhaires-air} used to generate the events. Also shown with dotted vertical lines are the average distances for the closest triggered antenna, second closest and so on, from left to right. In order to account for the intrinsic differences in the average EM energy of proton- and iron-induced showers, which would also affect the electric field amplitudes (see section~\ref{sec:zhairessims}),  on these LDFs we  normalized the electric field amplitudes by the EM energy of each shower. As a result, all showers now have the same EM energy, and any remaining amplitude differences cannot be attributed to variations in EM energy. Looking at the LDFs, one can see that the peak amplitude for each shower tends to increase as the position of the maximum moves closer to the shower core. In other words, the deeper a shower develops (the higher \xmax), the closer the peak is to the core and the larger its amplitude. This points to a very strong and well-behaved dependence of the peak amplitude on \xmax.

Using this LDF plot on the right panel of Fig.~\ref{fig:featureimportances}, we can now understand the feature importances of the random forest, shown on the left panel. In the region of the closest antenna (first vertical dotted line), we find the biggest amplitude differences between proton and iron showers, leading to the highest feature importance. The region of the second closest antenna (second vertical line) lies close to the point where the proton and iron LDFs tend to cross. Here, the amplitude differences between p and Fe are much smaller, explaining the sharp decrease in the feature importance for the second closest antenna. In the region of the third closest antenna, the amplitude differences increase again, gradually decreasing for antennas further away, matching the behavior of the feature importances on the left panel. So, it seems that the random forest is using the amplitude dependence on \xmax to perform the discrimination. Understanding this dependence motivates the present study, in which we revisit the radio LDF dependence on \xmax and explain, in a semi-quantitative way, why such a strong and well-behaved dependence emerges. Since the focus of this study is not the ML discrimination method itself, we defer a more detailed description of the method and its results to another work~\cite{MLpaper}.

It has been known for a long time that the shape of the radio LDF depends on composition. Earlier studies, such as those by the LOPES and Tunka-Rex collaborations, already exploited differences in the LDF between proton and iron showers to reconstruct $X_{\rm max}$, in particular through the slope of the radio LDF \cite{LOPESXmax,Tunka-E-Xmax,Tunka-E-Xmax-POS}. In these approaches, the composition dependence of the LDF arises primarily from geometrical effects related to the projection of the Cherenkov cone onto the ground. Showers that develop deeper in the atmosphere occur closer to the detector and therefore produce smaller radio footprints, leading to measurable differences in the LDF slope. In this sense, the shape of the LDF reflects the geometrical manifestation of the $X_{\rm max}$ dependence of the radio signal. While the dependence of the radio footprint shape on \xmax has long been recognized and exploited by such methods, the results presented here show that there is also an additional amplitude dependence related to the air density at \xmax and coherence effects.

With the introduction of the LOFAR \xmax reconstruction method~\cite{OriginalLofarXmax}, which is essentially a ``black-box-like'' $\chi^2$ comparison with multiple simulations, researchers stopped looking at LDF plots for multiple compositions.  In fact, to our knowledge, the last work in the literature to show LDF plots for multiple compositions was \cite{LOPESXmax}. Since the new LOFAR \xmax method performed much better than previous ones, it was widely adopted by cosmic ray radio experiments. And, given its ``black-box-like'' nature, understanding the LDF dependence on \xmax was no longer necessary. We believe this is the reason why this additional amplitude dependence on \xmax was historically overlooked, or attributed solely to differences in the EM energy of the showers.

\begin{figure}
  \begin{center}
    \includegraphics[width=0.5\linewidth]{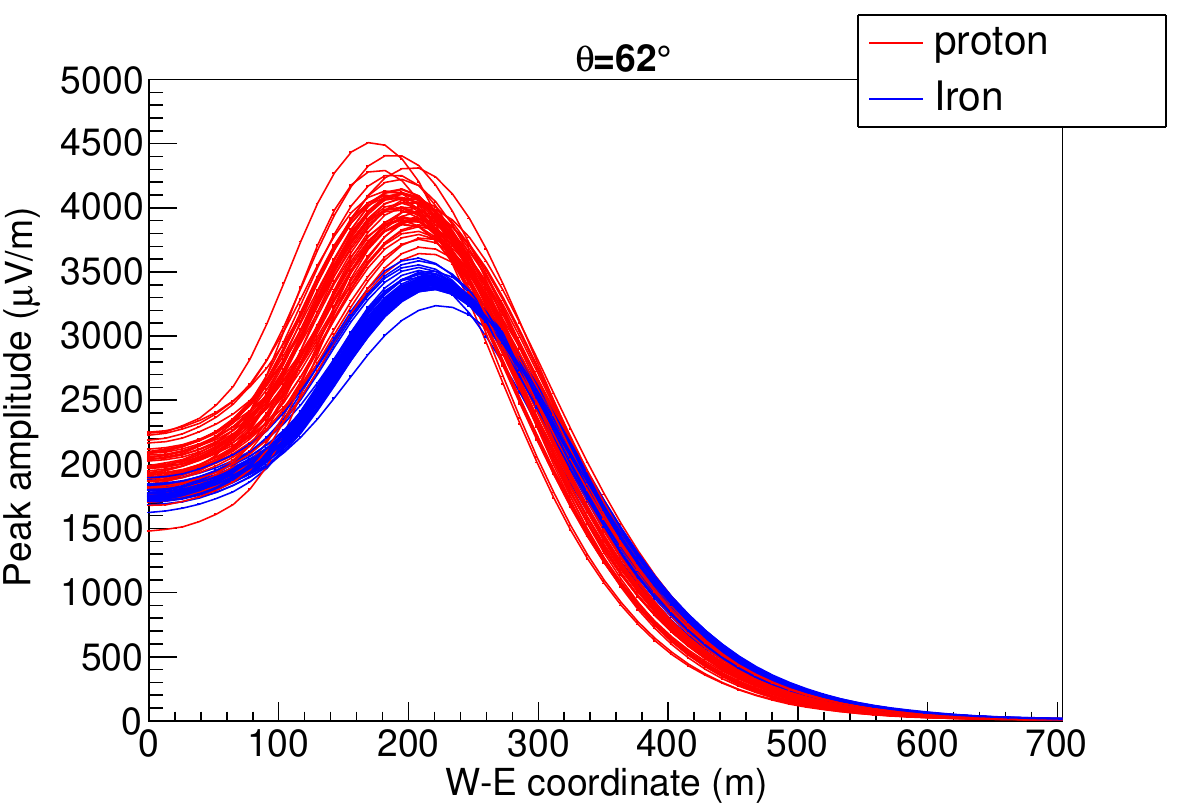}\includegraphics[width=0.5\linewidth]{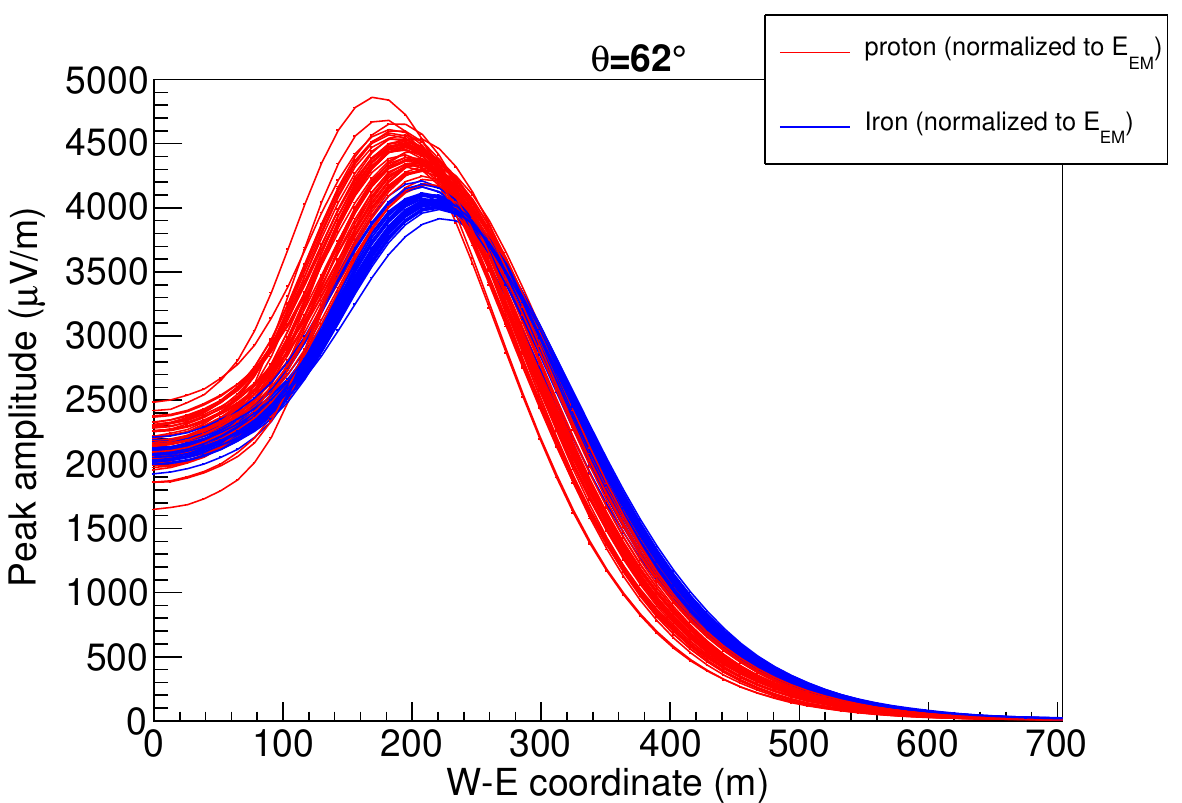}
    \caption{LDFs obtained from full ZHAireS simulations of events coming from the North with $\theta=62^\circ$ at GRAND. Left: Original LDF, without EM energy normalization (all showers have the same primary energy $E_0=1.25$ EeV). Proton-induced showers are shown in red and iron in blue. Right: Same as left, but the electric fields were normalized by the EM energy of each shower (all showers now have the same EM energy $E_{EM}=1.25$ EeV).}
    \label{fig:EMnormalization}
  \end{center}
\end{figure}

Since different primary masses lead to different \xmax distributions, an \xmax dependence also implies a composition dependence. Some plots in the literature, such as Fig. 6 of \cite{TimRadioRenaissance} and Fig. 1 of \cite{LOPESXmax} are consistent with this composition dependence on the LDF amplitudes. But, historically, any amplitude differences between showers induced by different primary compositions tended to be quickly dismissed as arising from the intrinsic differences in their EM energies, an interpretation that we show here is incomplete. This can be seen in Fig.~\ref{fig:EMnormalization}. On the left panel we show the original LDFs obtained from the full simulations (see section~\ref{sec:zhairessims}), without correcting for the different EM energy of each shower, i.e., all showers have the same primary energy but different EM energies. This means that the amplitude differences on the left panel are, to some extent, also due to differences in the EM energy of the showers. In contrast, on the right panel we normalized the LDF amplitudes by the EM energy of each shower, i.e., now all showers have the same EM energy and the remaining amplitude differences, although slightly smaller, cannot be attributed to different EM energies arising from the different primary compositions. In fact, one can now see a very well-behaved dependence of the amplitude on the position of the peak. The smaller the distance from the core to the peak, which relates to an increasing \xmax, the bigger the amplitude tends to be. This clearly shows we are dealing with an \xmax amplitude dependence. In this work we will show, in a semi-quantitative way, why this \xmax dependence arises and how it can be understood in terms of two simple scalings of the electric field. 

\section{Simulations of radio emission}
\label{sec:zhairessims}

In this work we used the \zhaires simulation package \cite{zhaires-air}, which stands for ZHS+AIRES, to model the radio emission of extensive air showers in the atmosphere. AIRES~\cite{aires} simulates the shower development, taking into account all its complexities, and provides the particle tracks that are used by the ZHS formalism~\cite{ZHS92,TimeDomainZHS} to calculate the net electric field at any observer position. \zhaires is based on first principles and does not a priori assume any emission mechanism. However, and as shown in \cite{zhaires-air}, the electric field obtained with \zhaires in the MHz-GHz frequency range is compatible with a superposition of the geomagnetic and charge excess radio emission mechanisms on most geometries.

Our simulations were performed for a single line of antennas at ground level, East of the shower core, using the SIBYLL hadronic model~\cite{SIBYLL}. This line contains $\sim$90 antennas, with a spacing that varies with zenith angle, so that the electric field can be calculated up to a distance of around 4.5 times the expected distance to the Cherenkov ring. For each zenith angle, a total of 50 proton- and 50 iron-induced showers were simulated at two sites: At the Giant Radio Array for Neutrino Detection (GRAND) and at the Pierre Auger Observatory (Auger). All simulations used the U.S. Standard Atmosphere 1976 (USatm76) model~\cite{USatm76}.

In the case of GRAND, we simulated showers with a primary energy ($E_0$) of 1.25 EeV, a ground altitude of 1264 m a.s.l. and a geomagnetic field of 56.4 $\mu T$, with an inclination of 61.6 deg. For this set, all showers come from the North with zenith angles varying from 42$^\circ$ to 82$^\circ$, in steps of 4$^\circ$. We also used a band-pass filter between 50 MHz and 300 MHz, to mimic the bandwidth of the GRAND detector. For the Auger site, we used $E_0=5$ EeV, a ground altitude of 1400 m a.s.l. and a magnetic field of 24.0 $\mu T$, with an inclination of -37$^\circ$. In this case, all showers come from the South with zenith angles varying between 55$^\circ$ and 85$^\circ$, in steps of 5$^\circ$, and the pulses were filtered between 30 MHz and 80 MHz.

For each antenna, the simulated electric field time trace was filtered in the corresponding detector bandwidth and the peak amplitude was defined as the maximum of the Hilbert envelope of the filtered time trace. These peak amplitudes were then used to construct the radio LDF as a function of distance to the shower axis. Later in this work, our analysis focuses on the maximum LDF amplitude, defined as the largest peak amplitude measured among all antennas in the LDF. This maximum LDF amplitude typically occurs close to the Cherenkov ring and represents the strongest radio signal measured at ground level for a given shower.

We estimated the EM energy of each shower by analyzing the EM particle stack in the simulations. The EM energy was defined as the sum of the deposited energy of all EM particles in the atmosphere, plus the energy of the EM particles discarded by the simulation during propagation, including low-energy ones. The average EM energies (missing/invisible energies) we obtained for each primary energy and composition are consistent with the ones in the literature, e.g., in~\cite{AugerInvisibleEnergy,ICRCBeijingMissingEnergy,MatiasMissingEnergy,KASCADEMissingEnergy}.

We also estimated the Geomagnetic-to-Askaryan ratio of each shower by disentangling the geomagnetic and Askaryan components of the emission, as described in \cite{toymodel}. This was done by repeating the simulations with the geomagnetic field set to 0. Note that this procedure was only used to estimate the Geomagnetic-to-Askaryan ratio and that all plots show the results of the simulations with $\vec{B}$ turned on.   

The goal of this work is to investigate the dependence of the electric field amplitude at ground on \xmax. Since the \xmax distribution depends on primary composition, the electric field amplitude is also composition dependent. It is also well known that the shower EM energy decreases with primary mass, meaning that in our simulations with a fixed $E_0$, proton-induced showers, on average, would tend to generate higher electric fields than iron-induced ones, even for showers with the same \xmax. To isolate the electric field dependence on \xmax, we normalized the simulated electric fields by the EM energy of each shower. By using this EM energy normalization, we isolate the amplitude composition dependence only to that arising from intrinsic \xmax differences. Also, since radio energy reconstruction methods measure EM energy~\cite{AERAEMEnergy,FelixTimInclinedRec,GlaserRadiationEnergy}, not primary energy, another advantage of this normalization is that our simulated events now mimic those that would be obtained in a single EM-energy bin from actual radio measurements.

\subsection{A few example radio LDFs}

\begin{figure}[htb]
  \begin{center}
    \includegraphics[width=0.333\linewidth]{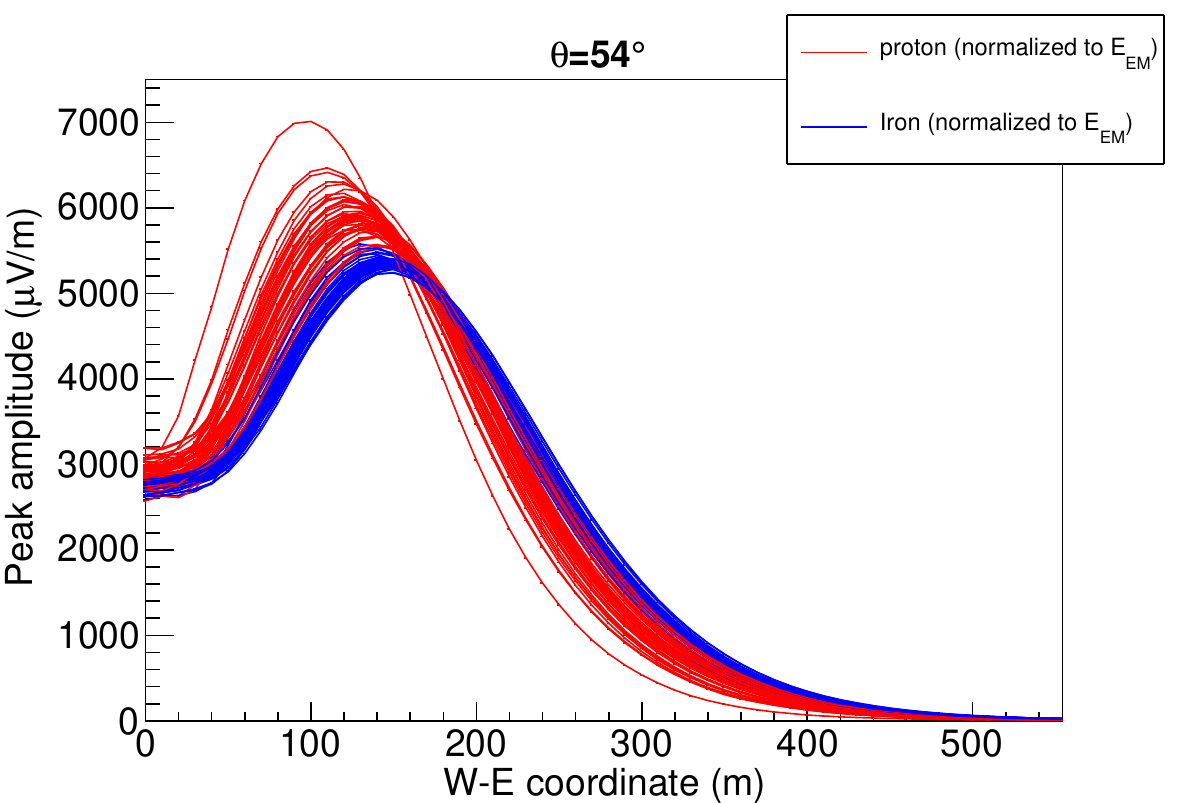}\includegraphics[width=0.333\linewidth]{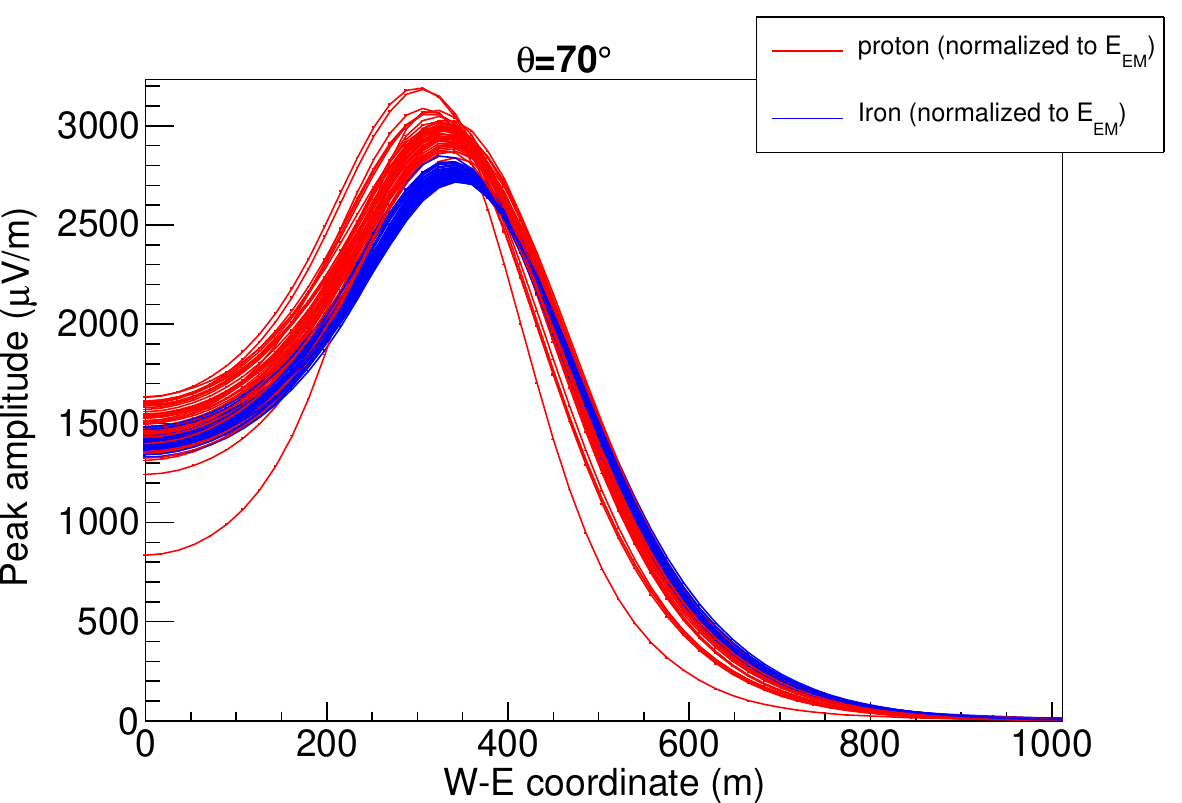}\includegraphics[width=0.333\linewidth]{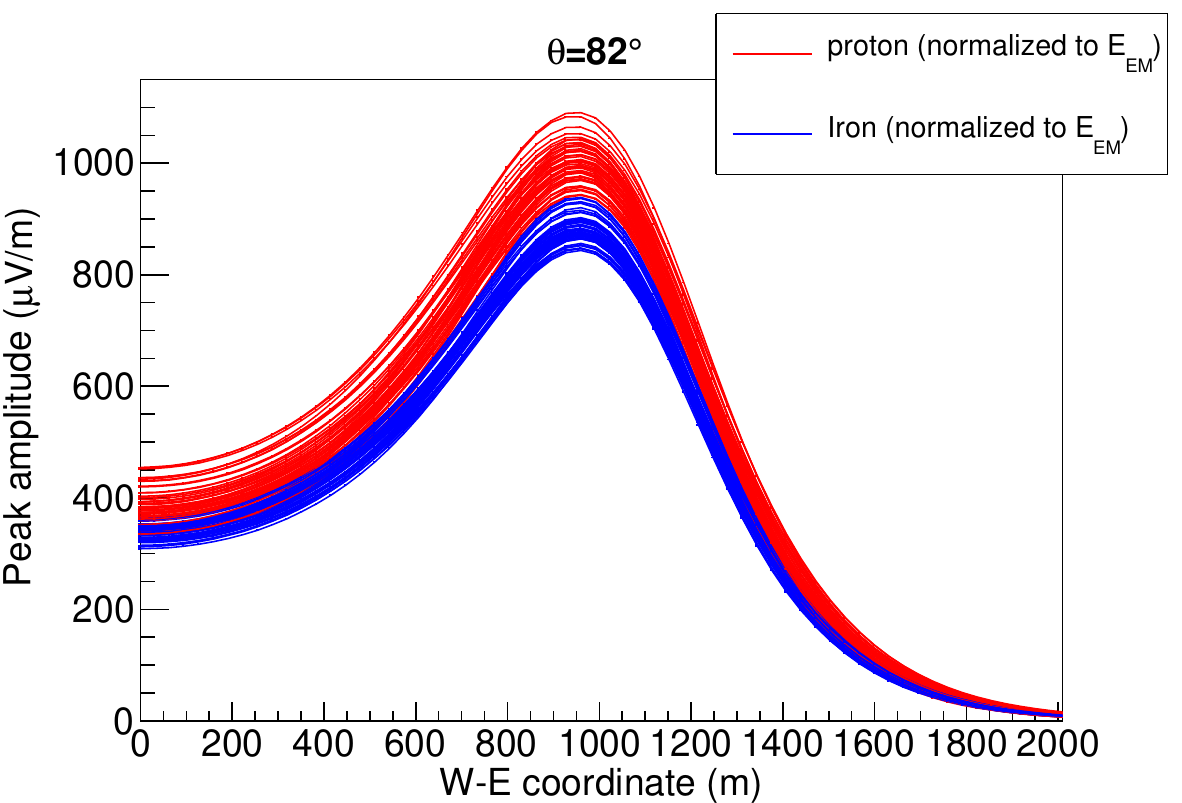}
    \includegraphics[width=0.333\linewidth]{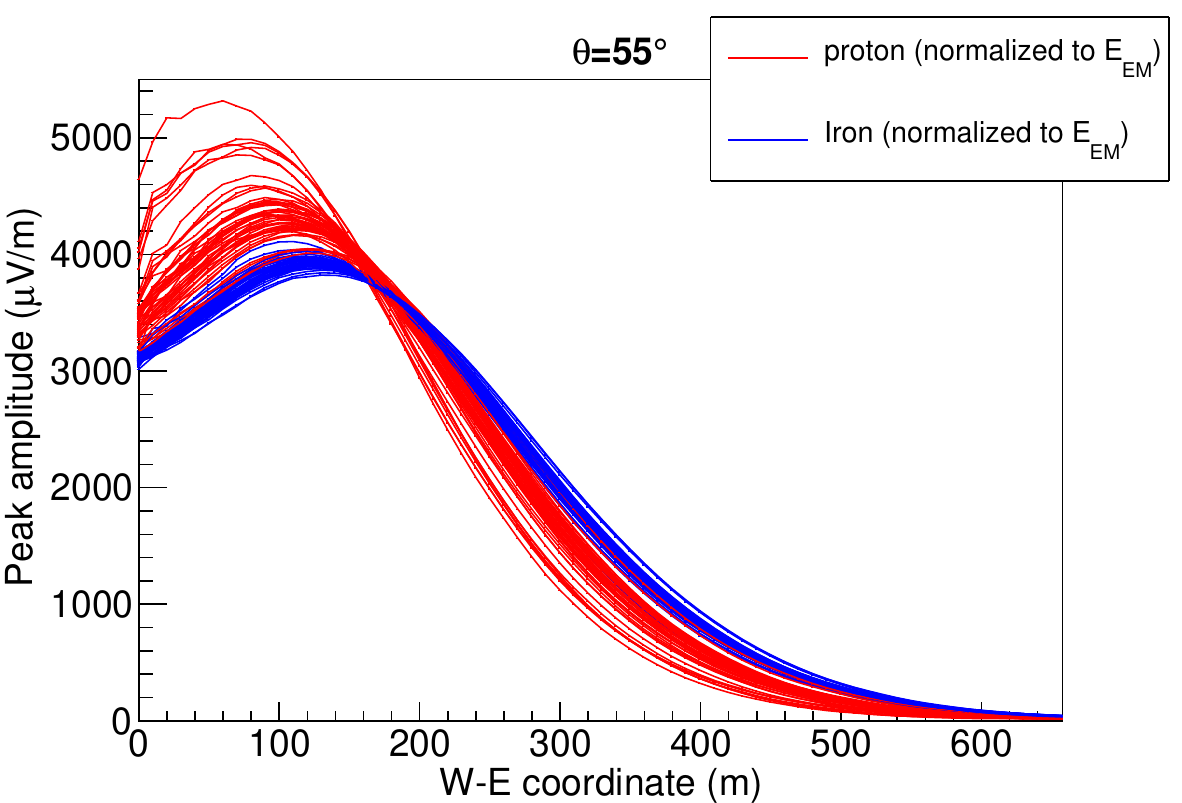}\includegraphics[width=0.333\linewidth]{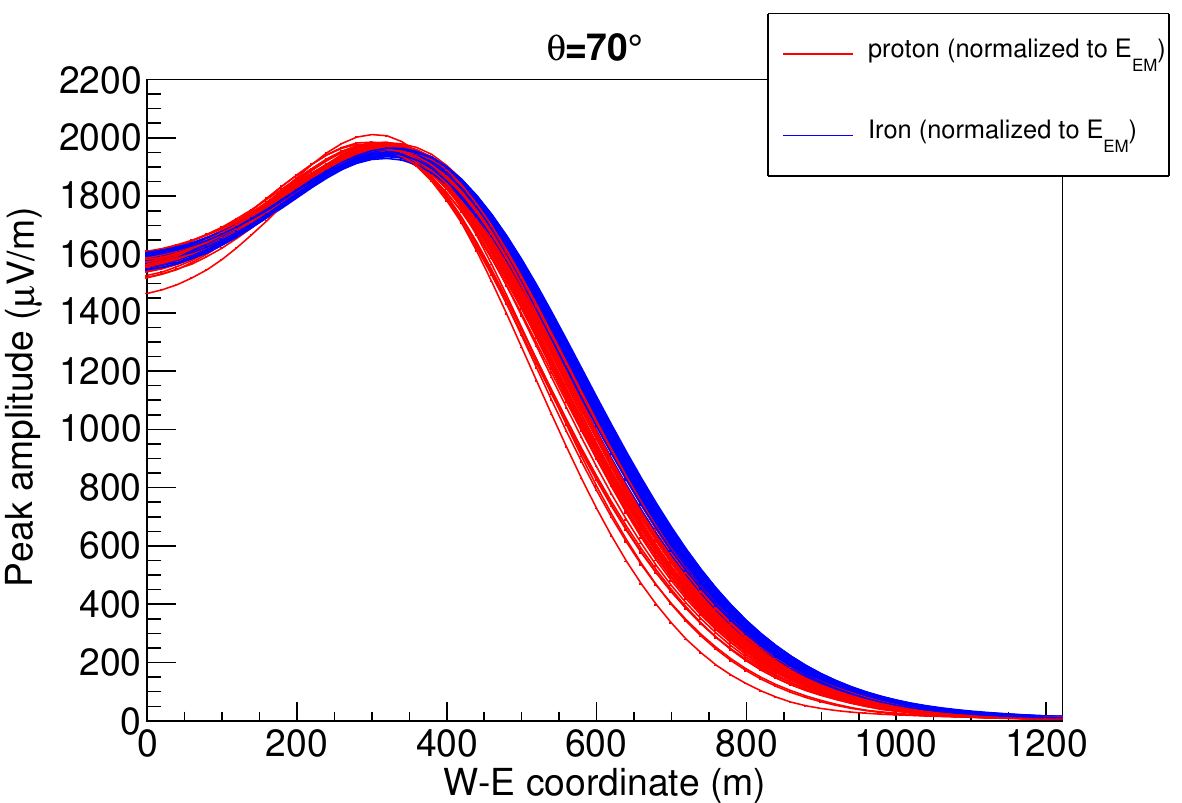}\includegraphics[width=0.333\linewidth]{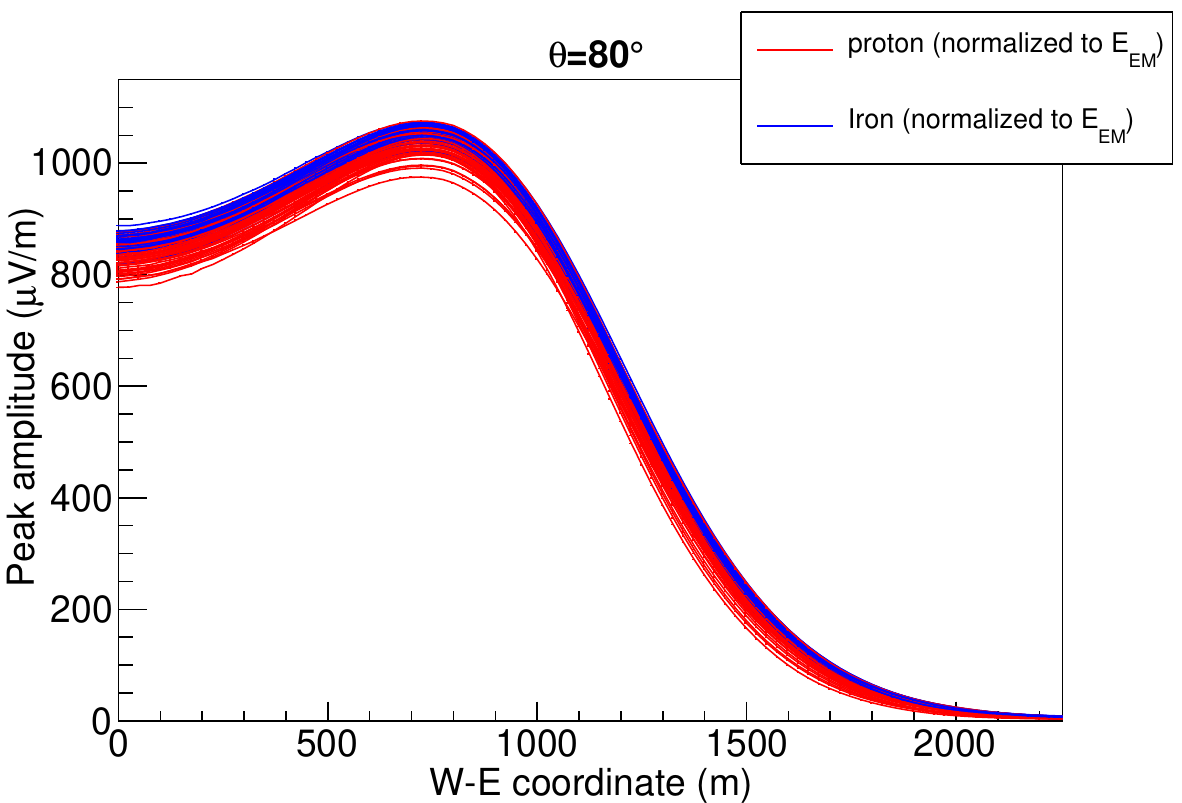}
    
    \caption{Radio LDFs from full ZHAireS simulations for several zenith angles at the GRAND site (top) and AUGER site (bottom). All LDFs were normalized by the EM energy of each shower. Figure extracted from~\cite{RLDF-ICRC2025}.}
    \label{fig:LDFExamples}
  \end{center}
\end{figure}

In Fig.~\ref{fig:LDFExamples}, we show examples of the radio LDF for a few zenith angles at the GRAND site (top) and at the Auger site (bottom). One can see that the shape of the LDF and the differences between proton and iron are strongly dependent on the zenith angle and detector site. At the GRAND site (top), the amplitude of proton-induced showers is, on average, higher than iron-induced ones at all studied zenith angles. In contrast, at Auger (bottom), proton showers tend to have higher fields at low zenith angles, while iron showers dominate at the highest zenith angles. The reason for this will become clear in the next sections.

It is also interesting to note that, at both sites, the shape of the radio LDF seems to be almost independent of \xmax at around 82$^\circ$, as the position of the Cherenkov ring does not change significantly between showers. At lower zenith angles, the radio footprint shape and size do vary with \xmax: As the position of \xmax gets higher in the atmosphere, the distance to the detector increases, which tends to increase the size of the radio footprint at ground level due to the projection effect. At the same time, the decrease in air density reduces the Cherenkov angle, tending to decrease the footprint size instead. The projection effect dominates at lower zenith angles, but at a zenith of around $82^\circ$, also referred to as ``the magic angle''~\cite{HarmMagicAngle}, the shower develops in a region of the atmosphere where these two competing effects approximately cancel each other, making the shape of the radio footprint practically independent of \xmax. However, one can see that although the shape does not change significantly, the amplitude still depends on \xmax. Since the exact value of the magic angle depends on atmospheric conditions and observation altitude, it should be regarded as a small angular range rather than a precisely defined angle.

\section{Radio emission in atmospheric showers}
\label{sec:radioemission}

At a macroscopic level, the radio emission of air showers can be thought of in terms of currents generated by the motion of charged particles. It can be shown \cite{TimeDomainZHS} that the induced electric field is approximately proportional to the projection of these currents along a direction perpendicular to the observation direction (perpendicular current $J_{\perp}$). At a microscopic level, this is the basis of the ZHS formalism \cite{ZHS92,TimeDomainZHS}, where the contribution of each particle track in the shower to the vector potential $\vec{A}(t)$ is given by:

\begin{equation}
  \vec{A}(t,\hat{u})=\frac{\mu e}{4\pi Rc^2} \vec{v}_{\perp}\frac{\Theta(t-t_1^{det})-\Theta(t-t_2^{det})}{1-n\vec{\beta}\cdot\hat{u}},\;\;\;\;\;\vec{E}(t)=-\frac{\partial{\vec{A}}}{\partial t},
  \label{eq:ZHS}
\end{equation}

\noindent where $c$ is the speed of light, $\mu$ is the permeability, $-e$ is the charge of an electron, $R\hat{u}$ is the vector from the midpoint of the track to the antenna, $n$ is the refractive index at the midpoint of the track, $\vec{v}$ is the speed of the particle inside the track (assumed constant), $\vec{\beta}=\vec{v}/c$, $\vec{v}_{\perp}=-[\hat{u}\times (\hat{u}\times\vec{v})]$ is the projection of the particle speed onto a plane perpendicular to the observing direction, $t$ is time at the detector, $t^{det}_{1}$ ($t^{det}_{2}$) is the arrival time at the detector of the signal coming from the beginning (end) of the particle track, $\Theta(x)$ is the Heaviside step function, and $\vec{E}(t)$ is the induced electric field at the antenna position (time trace). It is important to note that the ZHS formalism is based on first principles and does not presuppose any specific emission mechanism.  This is the formalism used by the ZHAireS radio emission simulation code\cite{zhaires-air}.

The radio emission of air showers is mainly due to the superposition of the geomagnetic and Askaryan (charge-excess) mechanisms. The geomagnetic emission arises from the opposite deflections of electrons and positrons in the geomagnetic field, producing a net current parallel to the Lorentz force. Since the electric field generated by the geomagnetic mechanism is proportional to the perpendicular current that arises from the Lorentz force, it depends on both the amplitude and orientation of the geomagnetic field, and is approximately proportional to $|\vec{B}|\sin(\alpha)$. The characteristic polarization of the electric field induced by the geomagnetic mechanism is largely independent of observer position and approximately parallel to $-\vec{V}\times\vec{B}$, where $\vec{V}$ is taken to be approximately parallel to the shower axis. Air density at the emission region also plays an important role in the geomagnetic emission, as will be discussed in detail later.

The Askaryan or charge-excess mechanism, as the name implies, arises from an excess of electrons in the shower. Since electrons and positrons both travel along the shower axis, the net current would be zero without an excess of electrons or positrons. But, as the shower develops, it entrains atomic electrons from the medium, creating an excess of electrons and a non-zero net current approximately parallel to the shower axis and proportional to the charge excess. The electric field generated by the Askaryan mechanism is polarized along the projection of this net current onto the plane perpendicular to the observer direction, and is thus approximately zero at the shower core, increasing with distance from it. This leads to an approximately radial polarization towards the shower axis \cite{ZHS92,toymodel}, with a strong dependence on observer position.

While the speed of charged particles along the shower axis is approximately constant and equal to the speed of light $c$, the perpendicular-to-axis speed, which comes mainly from the transverse momenta gained through interactions and the Lorentz force, is much smaller. Although this force tries to constantly increase the perpendicular momenta of the charged particles, there is a limit to the average perpendicular velocity, called the drift velocity \cite{scholten-driftvelocity}. This limit is due to the interactions of the charged particles with the molecules in the medium, which on average tend to randomize the transverse velocity. Since the geomagnetic radio emission arises from the current perpendicular to the shower axis, its intensity is then approximately inversely proportional to the air density at each stage of shower development. On the other hand, the component of the speed parallel to the shower axis is much larger and is unaffected by changes in air density, making the Askaryan contribution to the radio emission largely independent of air density.

The superposition of these two main emission mechanisms, with their different polarizations, creates the pattern (footprint) of the electric field at ground level. As the polarization of the Askaryan component depends on observer position, this pattern is asymmetric with respect to the shower core. However, since the polarization of the geomagnetic component is largely observer-independent, the larger the geomagnetic contribution, the less asymmetric the footprint becomes. Because the geomagnetic component is inversely proportional to air density, and inclined showers develop higher in the atmosphere, the footprint becomes more symmetric as the zenith angle increases~\cite{LOFAR_polarization}.

In Section~\ref{sec:Rrhoscaling} we further investigate how the radio emission depends on the air density at, and the distance to, the emission region, which we approximate as the position of \xmax in the atmosphere.

\section{Scaling of the electric field at ground level with distance and density }
\label{sec:Rrhoscaling}

It is well known that the characteristics of the radio footprint at ground level depend on \xmax. This dependence is the basis for many, if not all, \xmax reconstruction methods using the radio technique, such as the one developed by LOFAR~\cite{OriginalLofarXmax,LofarXmax2021,AERAXmaxPRL,AERAXmax}, which rely on empirical comparisons of the measured radio footprints with multiple simulations. While this procedure successfully exploits the dependence of the footprint on \xmax, the underlying physical scaling of the radio emission remains unclear, but could be understood in terms of basic principles: By inspecting Eq.~\ref{eq:ZHS}, one can see that the contribution of a particle track to the vector potential at the detector is proportional to $\vec{v}_\perp$ and inversely proportional to $R$. Since $\vec{v}_\perp$ scales with 1/$\rho$ \cite{scholten-driftvelocity}, we would then expect the electric field at the detector to scale with 1/$R$ and 1/$\rho$ at the emission point.

\begin{figure}[htb]
\begin{center}

  \includegraphics[width=0.5\textwidth]{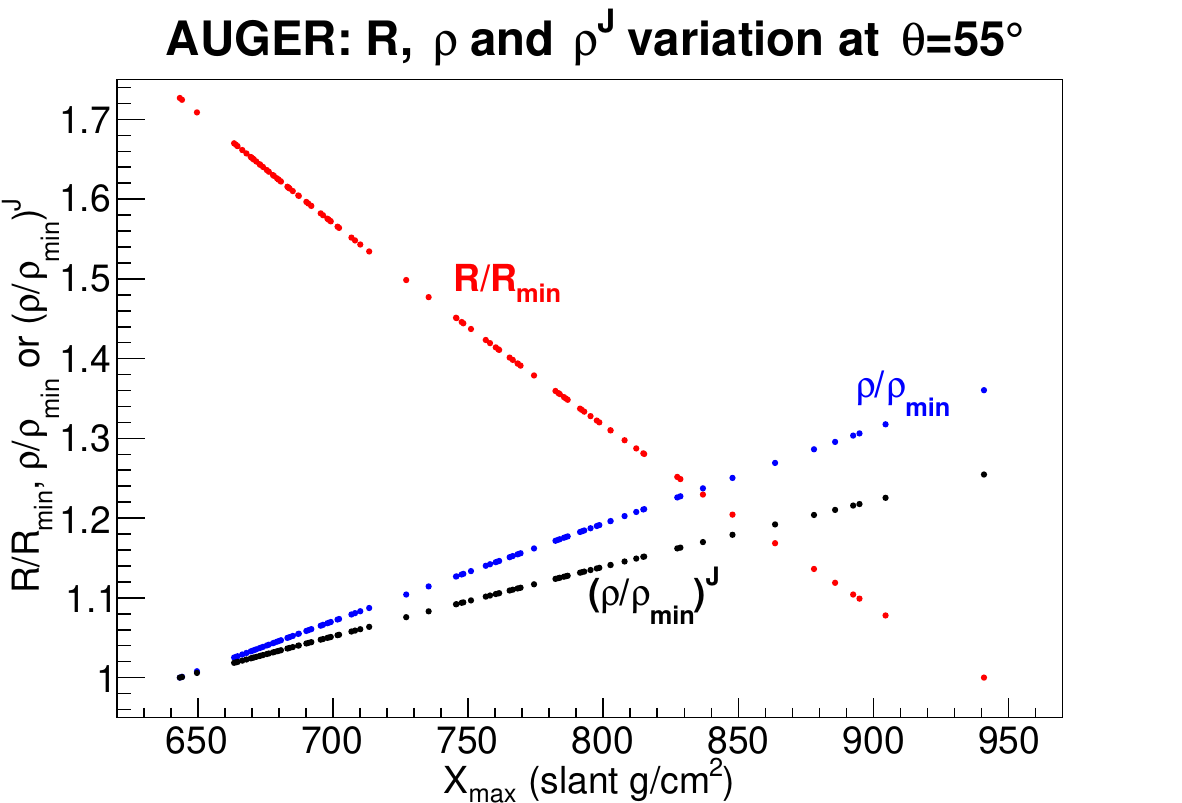}\includegraphics[width=0.5\textwidth]{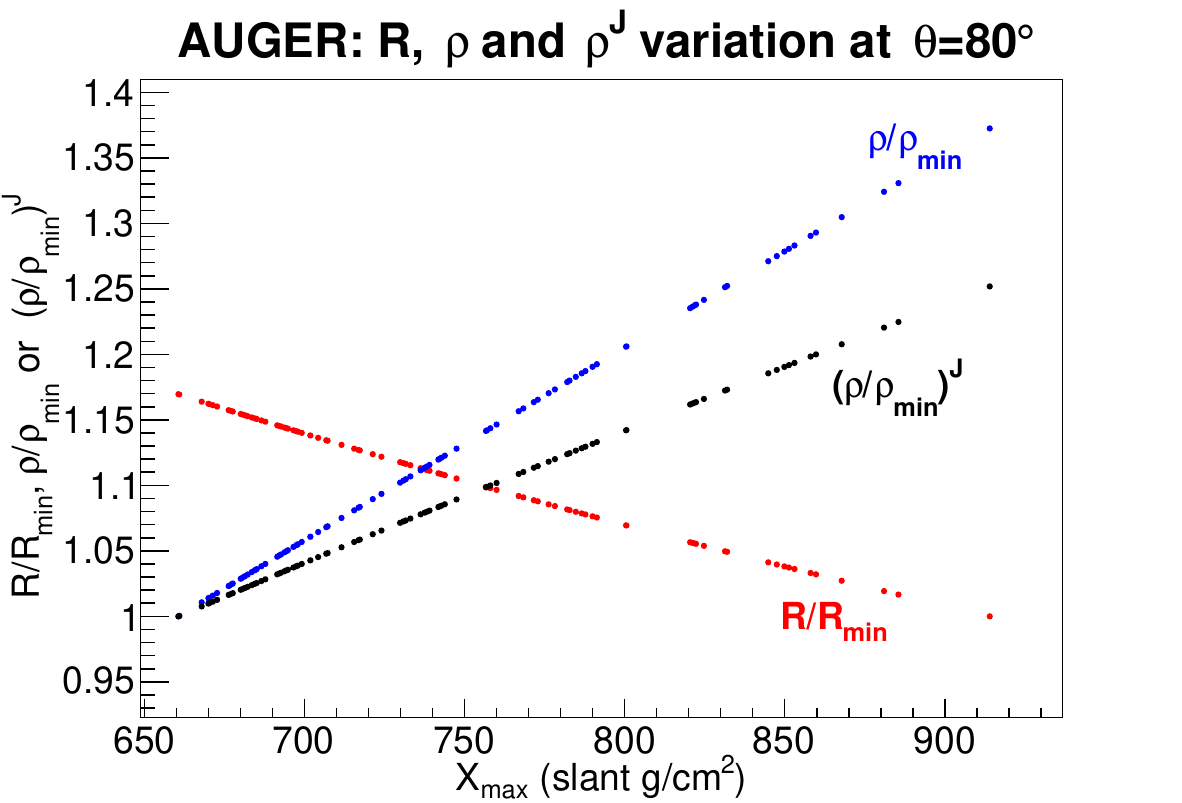}
  \includegraphics[width=0.5\textwidth]{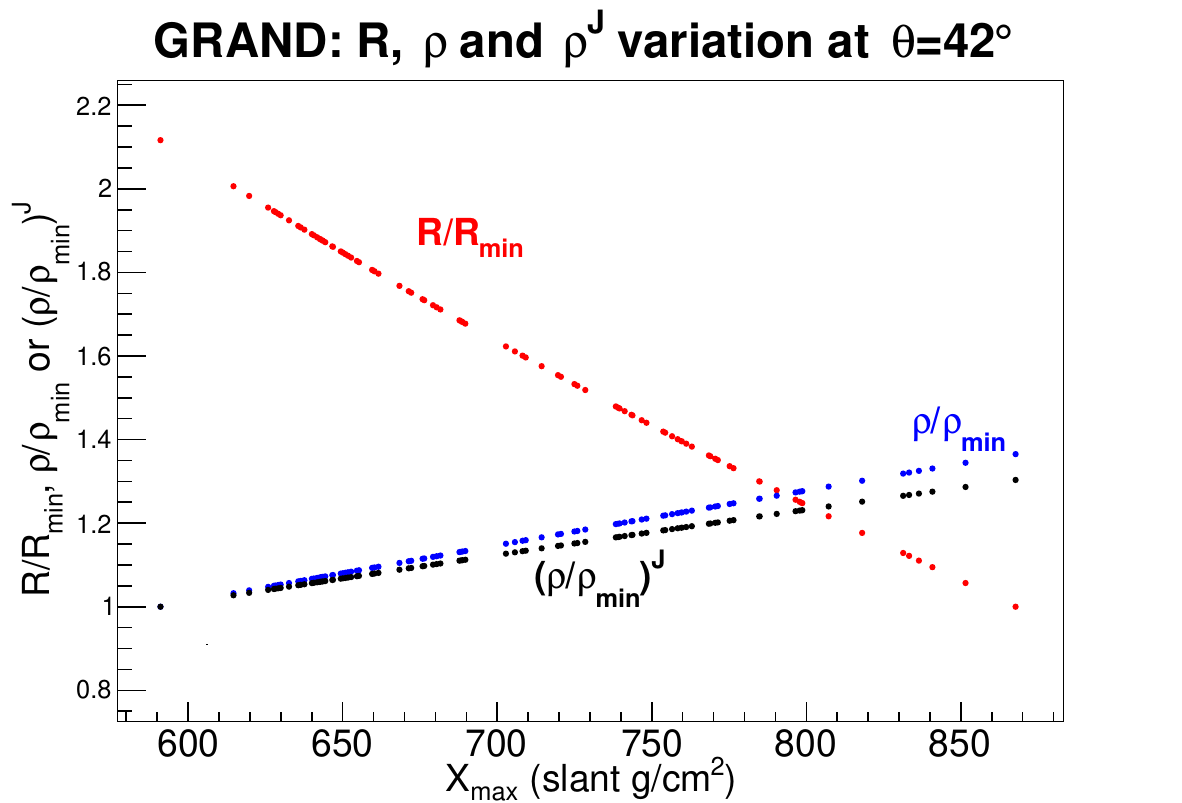}\includegraphics[width=0.5\textwidth]{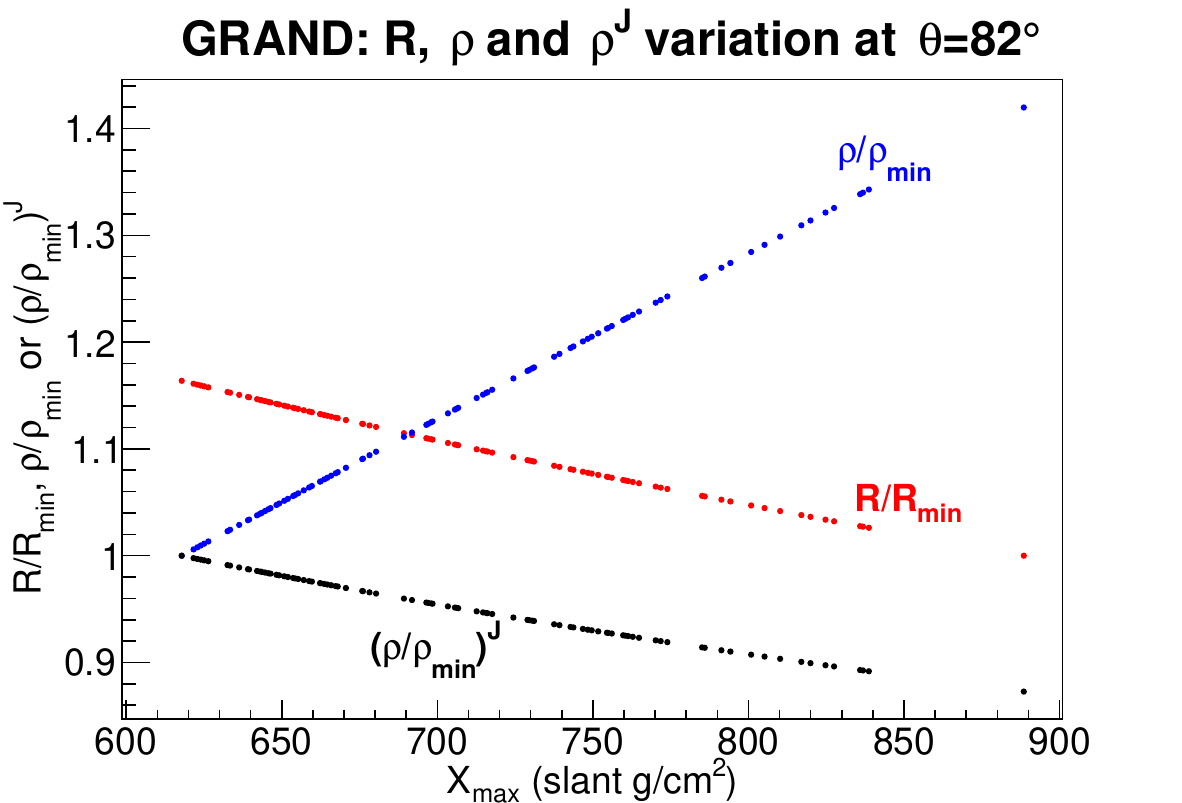}
\caption{Values of $R$ (red), $\rho$ (blue) and $\rho^{J(\theta)}$ (black) as a function of \xmax for several zenith angles and experimental sites, normalized by their respective minimum values for each zenith angle. $J(\theta)$ is a non-linearity factor used to take into account coherence loss (see section~\ref{sec:lossofcoherence}). Top left: Auger site  at $\theta=55^\circ$. Top right: Auger site  at $\theta=80^\circ$. Bottom left: GRAND site at $\theta=42^\circ$. Bottom right: GRAND site at $\theta=82^\circ$. Note that, except for the fitted $J(\theta)$ values (see section~\ref{sec:lossofcoherence}), these plots did not use the results of the simulations and instead were constructed solely using the shower geometry (zenith angle, ground altitude and \xmax) along with a model of the atmosphere (U.S. Standard Atmosphere 1976~\cite{USatm76}). Figure Extracted from~\cite{RLDF-ICRC2025}.}
\label{fig:RrhovsXmax}
\end{center}
\end{figure}

For the shower as a whole, it is very common to approximate the radio emission as if it were coming from the position of \xmax in the atmosphere. In this approximation, we define $R=R_{X_{max}}$ as the distance along the shower axis from the shower maximum to the ground and $\rho=\rho_{X_{max}}$ as the air density at the position of \xmax. Throughout the rest of this work, $R$ and $\rho$ refer to these values at \xmax. For a given shower geometry, the actual values of $R$ and $\rho$ will depend solely on a model of the atmosphere~\cite{USatm76} and on \xmax, i.e., $R=R(X_{max})$ and $\rho=\rho(X_{max})$ are monotonically decreasing and increasing functions of \xmax alone, respectively. This can be seen in Fig.~\ref{fig:RrhovsXmax}, where we show the values of $R$ (red) and $\rho$ (blue) as a function of \xmax for various zeniths and experimental sites, normalized by their respective minimum values for each given geometry.

As \xmax increases, showers develop closer to the detector, decreasing $R$ and increasing the electric field at ground level. On the other hand, as \xmax increases, the shower also develops at higher $\rho$. This decreases the drift velocity of the particles, and thus their perpendicular velocity $\vec{v}_{\perp}$ in Eq.~\ref{eq:ZHS}, ultimately reducing the electric field at ground level. So the 1/$R$ and 1/$\rho$ scalings are competing effects: 1/$R$ tends to increase the electric field emitted by deeper showers, while 1/$\rho$ tends to decrease it. It is well known that proton-induced showers tend to have a higher \xmax, on average, if compared to iron-induced ones. Therefore, this \xmax dependence also translates into a primary composition dependence: the 1/$R$ scaling will tend to increase the emission of proton-induced showers, while the 1/$\rho$ scaling will favor iron-induced ones.

Which of the two competing scalings will dominate depends on how much $R$ and $\rho$ vary with \xmax for a given shower geometry, which determines the region in the atmosphere where the shower develops. At low zenith angles, $R$ varies more than $\rho$ with \xmax, and the $1/R$ scaling dominates, on average favoring proton-induced showers. The opposite is true for high zenith angles, where $\rho$ varies more than $R$, making the $1/\rho$ scaling the dominant one, favoring iron-induced showers instead. This can be seen on Fig.~\ref{fig:RrhovsXmax}, where the values shown are normalized by their respective minimum values for each zenith angle. For Auger at 55$^\circ$ (GRAND at 42$^\circ$), $R$ varies by a factor of $\sim1.73$ ($\sim2.1$) and $\rho$ only varies by a factor of $\sim1.36$ ($\sim1.4$). At high zenith, for Auger at 80$^\circ$ (GRAND at 82$^\circ$), $R$ varies by a factor of $\sim1.17$ ($\sim1.16$) while the $\rho$ variation is larger, at $\sim1.37$ ($\sim1.42$). These same factors can be seen on Fig.~\ref{fig:FactorRatios}, where we show the relative variation of $R$ (red line) and $\rho$ (dashed line) for all zenith angles. Note that at this point we still disregard any loss of coherence, which increases as the air density decreases (see section~\ref{sec:lossofcoherence}), and assume that the $1/\rho$ scaling is valid for the whole atmosphere.

As discussed in Section~\ref{sec:motivation}, it has long been known that the shape of the radio LDF depends on \xmax, primarily through geometrical projection effects of the Cherenkov cone onto the ground. However, this geometrical effect represents only one manifestation of the \xmax dependence of the radio emission. The absolute amplitude of the radio signal also depends on the interplay between the distance and density scalings described above, and thus on \xmax. This becomes particularly clear near the so-called ``magic angle'', where the LDF shapes become nearly identical for different showers while the peak amplitude retains a clear dependence on \xmax. This behavior, illustrated in the right panels of Fig.~\ref{fig:LDFExamples}, reveals an additional physical component of the \xmax dependence of the radio signal.

Since the peak amplitude varies with the position along the LDF, for simplicity we will base all following analyses on the peak amplitude at a single point, namely the maximum LDF amplitude near the Cherenkov ring. In the subsequent analysis, references to peak amplitudes or electric field amplitudes refer to this maximum LDF amplitude. In order to approximate the emission of each mechanism, as it would appear at \xmax, we used our estimate of the geomagnetic-to-Askaryan ratio of each shower (see section~\ref{sec:zhairessims}) to obtain the maximum LDF amplitude at ground for the Geomagnetic (Geo) and Askaryan (Ask) components separately. We then multiplied Ask and Geo by the distance $R$ to the maximum of each shower, i.e., we applied the inverse distance scaling. Since the geomagnetic emission is also proportional to $\sin(\alpha)$, and $\alpha$ changes with zenith angle, we have also divided the field Geo at ground by $\sin(\alpha)$, in order to remove any zenith angle dependence.

\begin{figure}[htb]
  \begin{center}
    \includegraphics[width=0.5\textwidth]{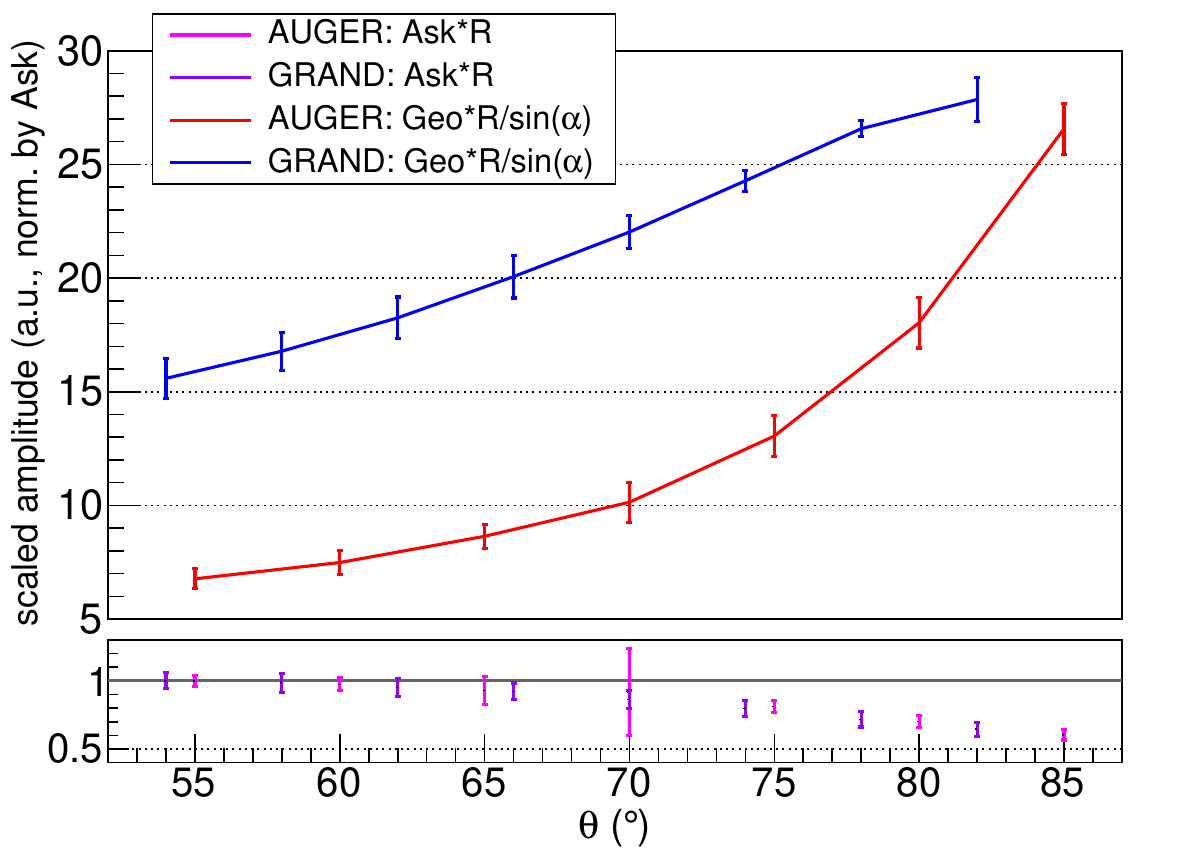}\includegraphics[width=0.5\textwidth]{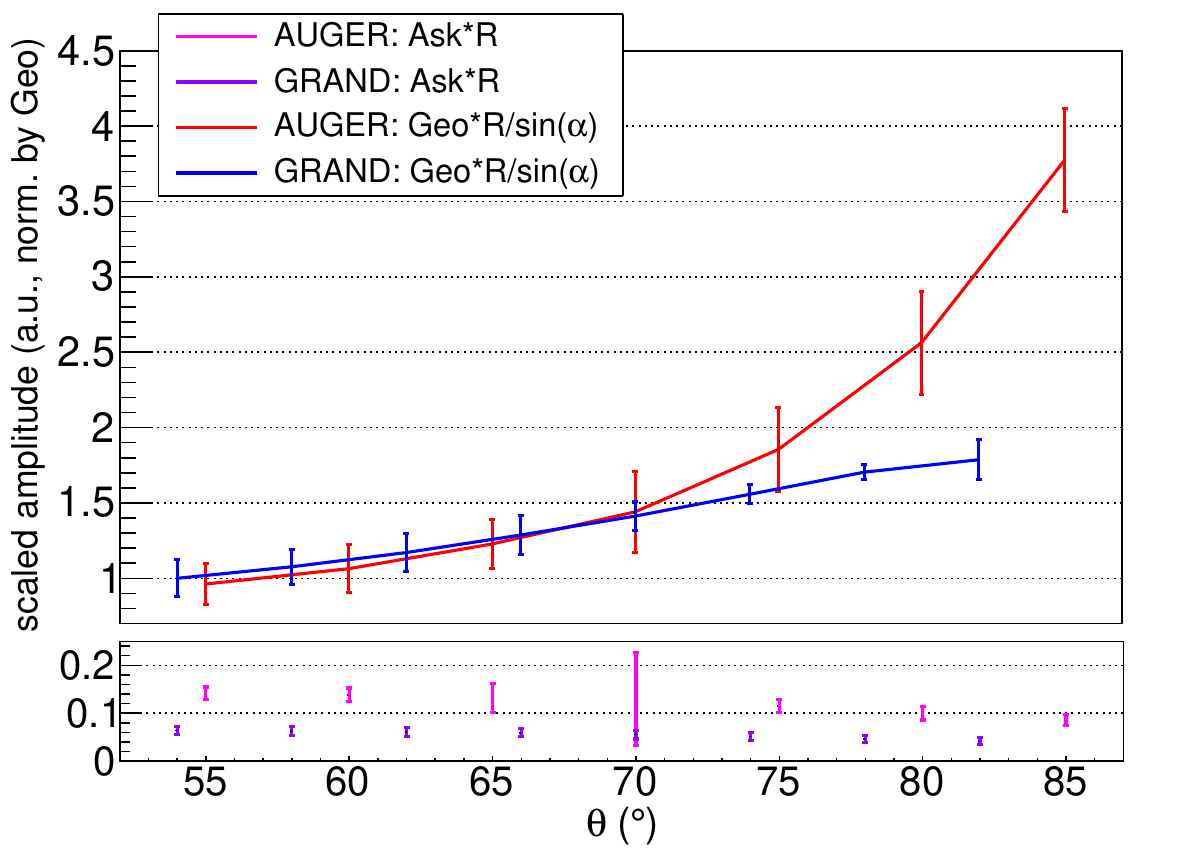}
    \includegraphics[width=0.5\textwidth]{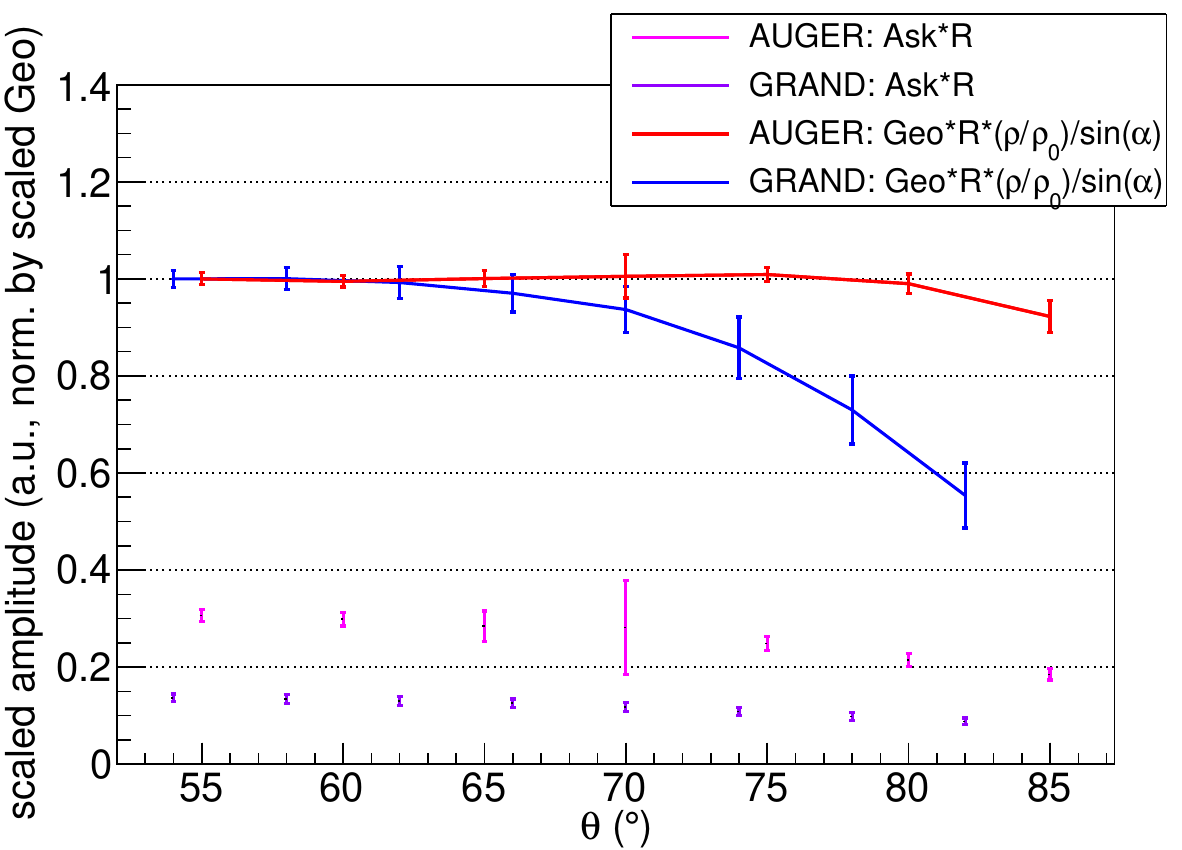}\includegraphics[width=0.5\textwidth]{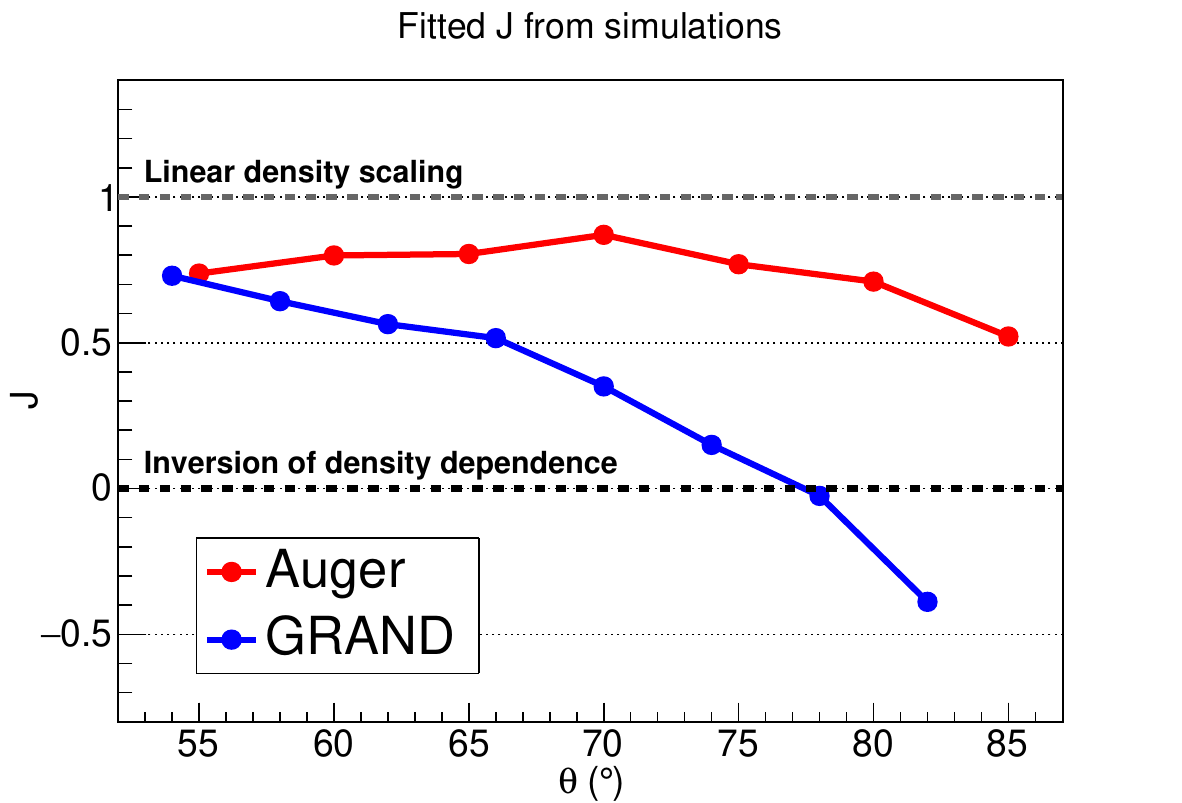}
    \caption{Top left: Estimate of the Askaryan and geomagnetic emission at GRAND and Auger, obtained from the complete set of full ZHAireS simulations. In this panel, the values are normalized by the Askaryan amplitude. Top right: Same as top left, but normalized by the geomagnetic emission. Bottom left: Askaryan and geomagnetic amplitudes after applying the inverse distance and density scalings, normalized independently for each site to its scaled geomagnetic amplitude at the lowest zenith angle. Bottom right: Non-linearity factor $J(\theta)$, which represents the coherence loss, fitted to the simulations for each zenith angle. The horizontal grey line indicates $J=1$, corresponding to the linear density scaling assumed in earlier analyses. The zenith bins differ between Auger and GRAND, reflecting the independent simulation sets performed for each site. For more details on all panels, see the text.}
    \label{fig:Geo-Ask-emission}
  \end{center}
\end{figure}

On the top panels of Fig.~\ref{fig:Geo-Ask-emission}, we show estimates of the Geomagnetic and Askaryan emission as a function of zenith angle for our whole simulated dataset, at the Auger and GRAND sites. The most relevant difference between the sites is that the geomagnetic field at GRAND ($|\vec{B}|=56.4\,\mu\mathrm{T}$) is more than double that at Auger ($|\vec{B}|=24.0\,\mu\mathrm{T}$). Both panels show the same exact results, but are just normalized differently to highlight different characteristics of the field dependence on zenith angle. In the top left panel we normalized the emission by the Askaryan emission, while on the top right panel we show the same data, but normalized by the geomagnetic emission instead. On the top left panel of Fig.~\ref{fig:Geo-Ask-emission}, one can see that the Askaryan emission exhibits a comparatively weak zenith dependence, showing a gradual decrease toward larger zenith angles, possibly reflecting geometrical and density effects associated with shower development at lower air densities. Nevertheless, this variation remains much smaller than the intrinsic zenith evolution of the geomagnetic component, most clearly visible in the Auger geomagnetic curve, and indicates that the $1/R$ scaling provides a good first-order description across the whole atmosphere. Also, one can see that the geomagnetic emission at GRAND, especially at low zenith angles, is much higher than that at Auger. At $\theta\sim55^\circ$, the Geomagnetic emission at Auger is only $\sim6.5$ times bigger than the Askaryan emission, while at GRAND it is $\sim15.5$ times bigger. This difference is consistent with the increase by a factor of $\sim2.3$ of the GRAND geomagnetic field, compared to the Auger one. As we increase the zenith angle, the shower develops in a more rarefied atmosphere, increasing the perpendicular current in Eq.~\ref{eq:ZHS} and the geomagnetic emission. In the same figure, one can clearly see that the geomagnetic emission increases with zenith both at Auger and GRAND. But it increases faster at Auger, reaching a similar emission to that of GRAND at $\sim85^\circ$, despite the much lower geomagnetic field\footnote{This may be relevant for experiments aiming to measure only very inclined showers. Traditionally, one of the most important factors when choosing a radio detector site is the geomagnetic field in the region, as in general a high geomagnetic field tends to increase the radio emission significantly. But in the case of high zenith angles, the geomagnetic field has only a small net impact on the measured field amplitudes, due to the higher loss of coherence related to higher geomagnetic fields. This can be seen on the top left panel of Fig.~\ref{fig:Geo-Ask-emission}. So, choosing a site with a low geomagnetic field becomes an option for experiments focusing on almost horizontal showers.}.
On the top right panel of Fig.~\ref{fig:Geo-Ask-emission}, normalized by the geomagnetic emission, one can see that the geomagnetic emission over the whole zenith range increases by a factor of $\sim3.75$ at Auger, compared to the factor of only $\sim1.75$ at GRAND. The stronger geomagnetic field at GRAND increases the loss of coherence, which in turn slows the increase of the geomagnetic emission with zenith. We discuss this loss of coherence, and its relation to the geomagnetic field, in section~\ref{sec:lossofcoherence}.

\subsection{Loss of coherence}
\label{sec:lossofcoherence}

Recently, several studies have reported a loss of coherence of the radio emission at low air densities, especially in the presence of a high geomagnetic field~\cite{JuanLossCoherence,ChicheLossCoherence,GuelfandLossCoherence}. As air density decreases, the drift velocity increases, which in turn increases the perpendicular current and the geomagnetic emission (see Eq.~\ref{eq:ZHS}). On the other hand, the larger deflections at lower densities increase the time delays between the tracks in the shower, and reduce the coherence of the geomagnetic radio emission. An increase in the geomagnetic field would lead to similar effects, due to the increase in the strength of the Lorentz force (see section~\ref{sec:radioemission}). This means that as we decrease the air density or increase $B$ we have two competing effects: An increase in the perpendicular current, which would tend to increase the geomagnetic emission, and an increase in the time delays, which decreases coherence and would tend to decrease the emission instead. The perpendicular current increase dominates in most cases, especially at higher air densities or lower magnetic fields, leading to a net increase in the geomagnetic emission as the air density decreases.

This loss of coherence has repercussions on the $1/\rho$ scaling we initially proposed, which is solely based on Eq.~\ref{eq:ZHS} and disregards any coherence loss. One would expect that as the air density at \xmax decreases, e.g., by increasing the zenith angle, the increase in $J_{\perp}$ would be partially counterbalanced by the accompanying increase in coherence loss. This means that the simple $1/\rho$ scaling cannot be valid for the whole atmosphere.

One can visualize the expected decrease in the strength of the density scaling by looking at the bottom left panel of Fig.~\ref{fig:Geo-Ask-emission}, where the full inverse scaling was applied to the geomagnetic emission. This was accomplished by multiplying Geo by both $R$ and $(\rho/\rho_0)$, where $\rho_0$ is the air density at sea level. This inversely scaled geomagnetic amplitude should remain approximately constant while the original $1/\rho$ scaling, which does not account for the loss of coherence, is approximately valid. One can see that at Auger, with its lower $|\vec{B}|$, the inversely scaled amplitude is somewhat constant up to $\sim80^\circ$, when it starts to decrease. But at GRAND, where the geomagnetic field is more than twice as strong, the scaling becomes non-linear much earlier, at around $\sim66^\circ$. This highlights the strong dependence of coherence loss on $|\vec{B}|$.

To account for this loss of coherence, we introduced a non-linearity factor $J(\theta)$ into the density scaling, which now becomes $(1/\rho)^{J(\theta)}$. This factor $J(\theta)$ was fitted to the simulations. For each zenith angle separately, we minimized the dispersion of the geomagnetic peak amplitudes when the full inverse scaling was applied, i.e., we minimized the dispersion of Geo$R(\rho/\rho_0)^{J}$. In other words, we fitted the strength of the density scaling to the simulated geomagnetic fields at each zenith angle, effectively estimating the average loss of coherence.

In the bottom right panel of Fig.~\ref{fig:Geo-Ask-emission}, we show the fitted values of $J(\theta)$ as a function of zenith for the simulations at the GRAND site. Note that $J(\theta)$ becomes negative for $\theta \gtrsim 77^\circ$. Above this zenith, the combination of low density and the large geomagnetic field at GRAND makes loss of coherence the dominant effect. This reverses the scaling of the geomagnetic emission amplitude with density at high zeniths, due to coherence loss outweighing the increase in perpendicular current.

On Fig.~\ref{fig:RrhovsXmax}, we also show the variation of $\rho^J$ (black) as a function of $X_\text{max}$, normalized by its minimum value, along with the previously discussed $R$ (red) and $\rho$ (blue) values. One can see that, in general, the introduction of the $J$ factor decreases the variation of $\rho^J$, which in turn decreases the strength of the density scaling, if compared to the $1/\rho$ linear case. At Auger, the much smaller geomagnetic field is never high enough to reverse the $\rho$ scaling, even at $\theta=80^\circ$ (top right panel of Fig.~\ref{fig:RrhovsXmax}). On the other hand, for GRAND at $\theta=82^\circ$ (bottom right panel of Fig.~\ref{fig:RrhovsXmax}), the reversal of the density scaling, due to the much higher geomagnetic field, can be clearly seen.

\section{Predictions based on the proposed scalings and comparison to the simulation results}
\label{sec:PredictionsAndComparison}

In this section we will make predictions for the dependence of the electric field amplitudes on \xmax and then compare them to the results of our full simulations for different zeniths and sites. In Sec.~\ref{sec:Rrhoscaling} we considered the basic competing $1/R$ and $1/\rho$ scalings. By using only the shower geometry and an atmospheric model, we established the basic geometrical behavior of these scalings, without taking coherence loss into account: the distance scaling dominates at low zeniths, favoring the deeper proton-induced showers, while the density scaling dominates at high zeniths, favoring the shallower iron-induced ones. In Sec.~\ref{sec:lossofcoherence}, we discussed the coherence loss effect, which reduces the strength of the density scaling. To account for this, we introduced a non-linearity factor into the density scaling, which then became $(1/\rho)^{J(\theta)}$. In addition, as described in Section~\ref{sec:zhairessims}, all showers are normalized by their individual EM shower energy obtained from the simulations. This means that all showers, regardless of primary composition, have the same EM energy, and any amplitude variations cannot be attributed to differences in EM energy.

\begin{figure}[htb]
  \begin{center}
    \includegraphics[width=0.5\textwidth]{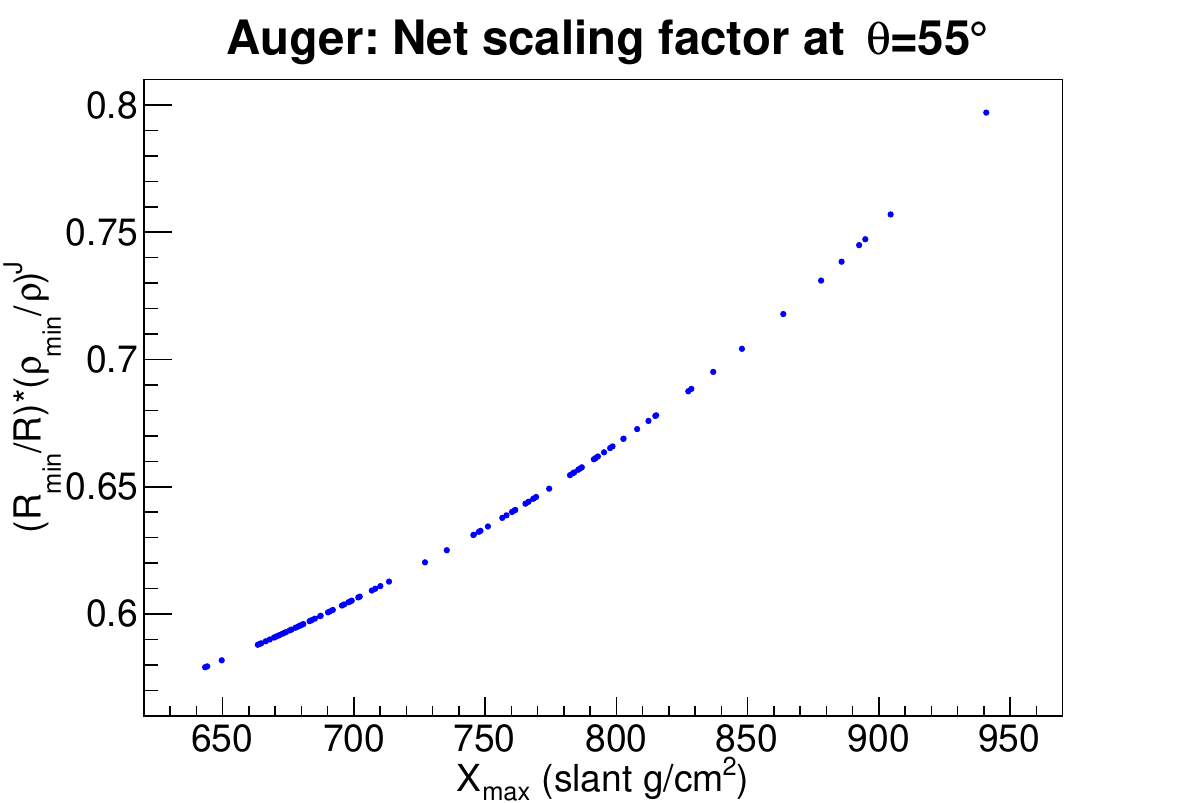}\includegraphics[width=0.5\textwidth]{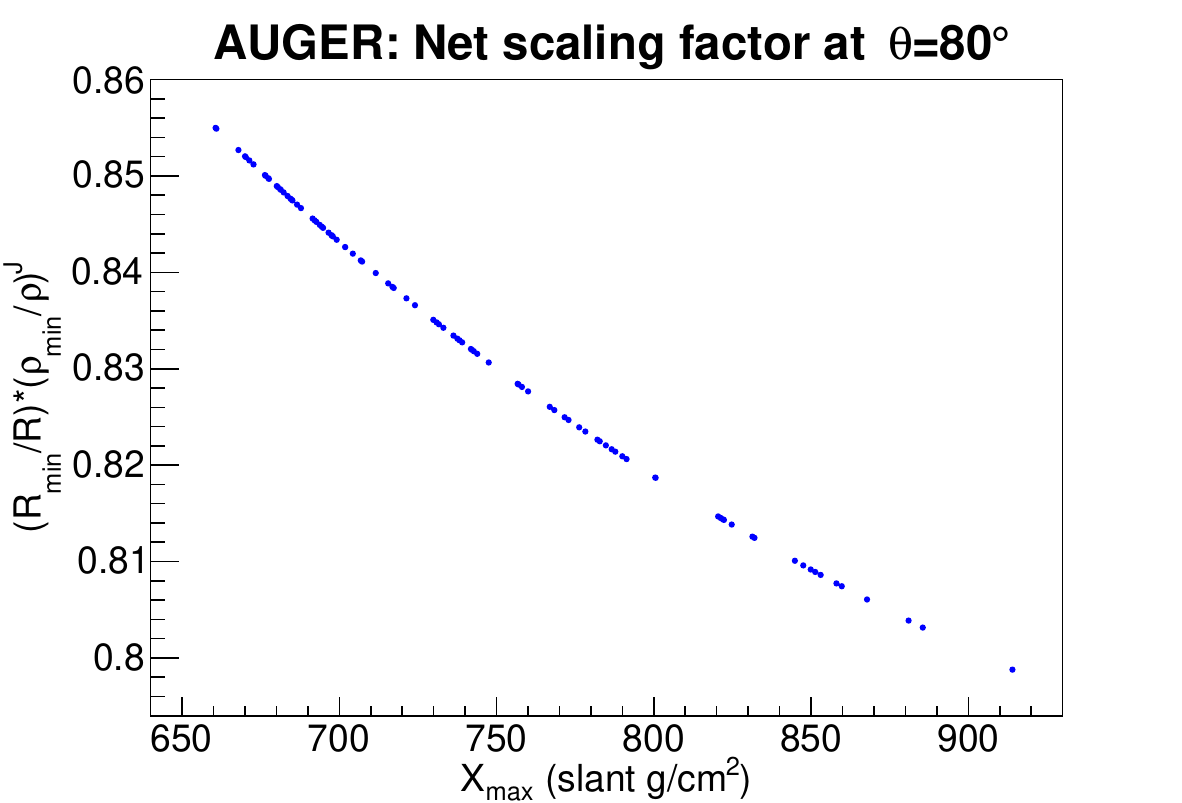}
    \includegraphics[width=0.5\textwidth]{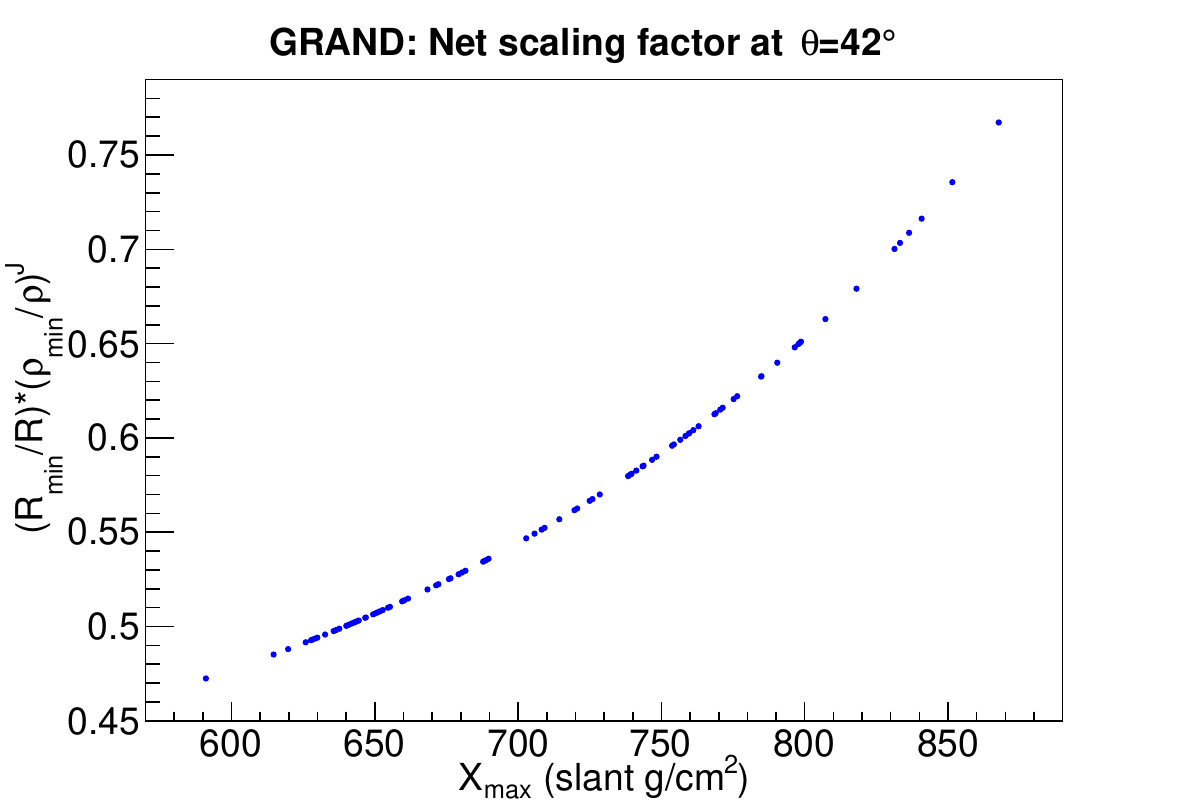}\includegraphics[width=0.5\textwidth]{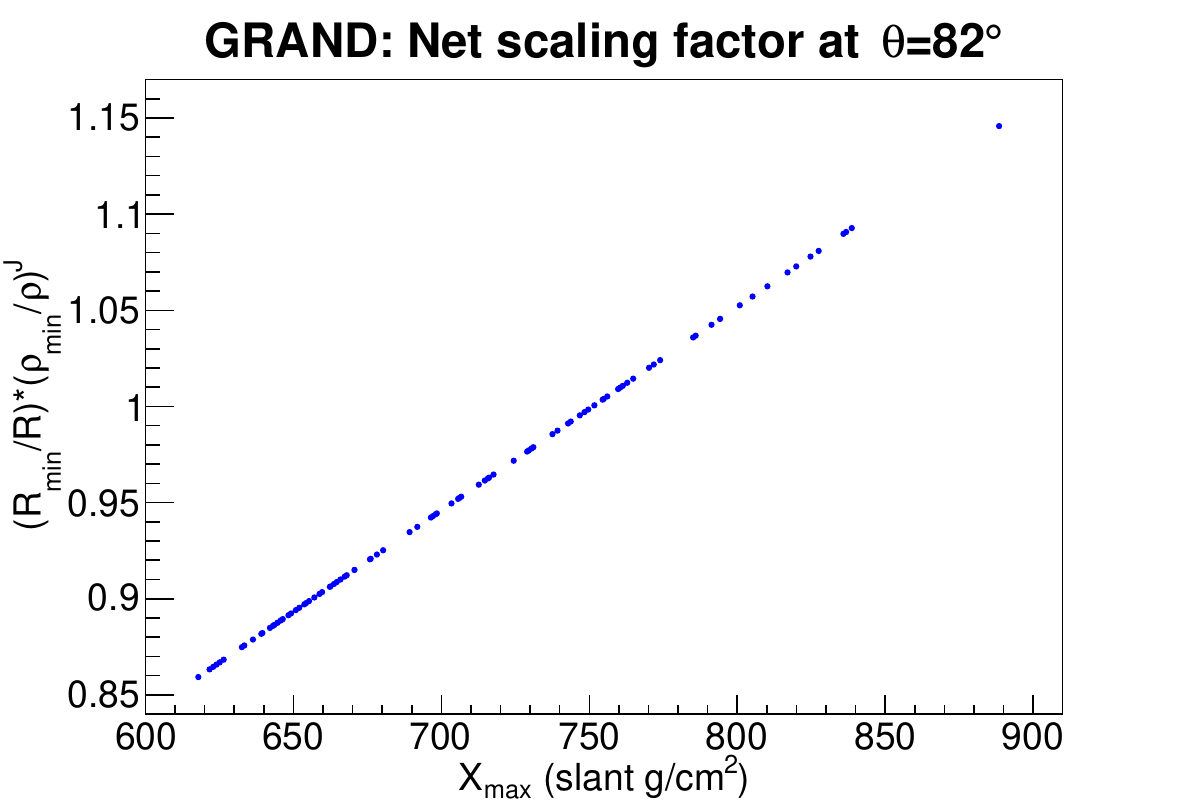}
    \caption{Net scaling factor $(R_{min}/R)(\rho_{min}/\rho)^{J}$ for the Auger site (top) and GRAND (bottom) for lower zenith angles (left) and higher zenith angles (right). These represent the expected change of the amplitudes with \xmax, by using the proposed density and distance scalings together. See text for more details.}
    \label{fig:NetScaling}
  \end{center}
\end{figure}

To visualize the combined effect of the $(1/R)$ and $(1/\rho)^{J(\theta)}$ scalings, we define the net scaling as $(R_{\rm min}/R)\allowbreak(\rho_{\rm min}/\rho)^{J(\theta)}$, shown in Fig.~\ref{fig:NetScaling} as a function of \xmax. This figure was constructed using the previously calculated variations of $R$ and $\rho^{J(\theta)}$ (Fig.~\ref{fig:RrhovsXmax}) for the same zeniths and detection sites. On the top panels of Fig.~\ref{fig:NetScaling}, one can see that the net scaling factor at the Auger site increases with \xmax at low zenith angles (top left panel), but decreases at high zeniths (top right panel). This behavior follows the geometrical expectation, as at Auger the geomagnetic field and loss of coherence are not strong enough to reverse this trend. On the other hand, at the GRAND site, the net scaling factor increases with \xmax both at low and high zeniths (bottom panels of Fig.~\ref{fig:NetScaling}). The high loss of coherence at GRAND, due to its very high $\vec{B}$, decreases the strength of the density scaling to such a degree that the distance scaling remains dominant even at large zeniths.

In order to investigate which of the two proposed scalings dominates at each zenith angle, $1/R$ or $(1/\rho)^{J(\theta)}$, we have used the relative variations of $R$ and $\rho^{J(\theta)}$ shown in Fig.~\ref{fig:RrhovsXmax}. As can be seen in the figure, the relative variations of $R$ and $\rho^{J(\theta)}$ are approximately linear with \xmax. So, the differences between the minimum and maximum values of each are approximately proportional to their derivatives with respect to \xmax. We then took the ratios $R_{Ratio}(\theta)=[R_{max}(\theta)/R_{min}(\theta)]$ and $\rho_{Ratio}^J(\theta)=[\rho_{max}(\theta)/\rho_{min}(\theta)]^{J(\theta)}$ at each zenith angle as a measure of the relative strength of the $1/R$ and $(1/\rho)^{J(\theta)}$ factors, respectively.

\begin{figure}[htb]
\begin{center}

  \includegraphics[width=0.5\textwidth]{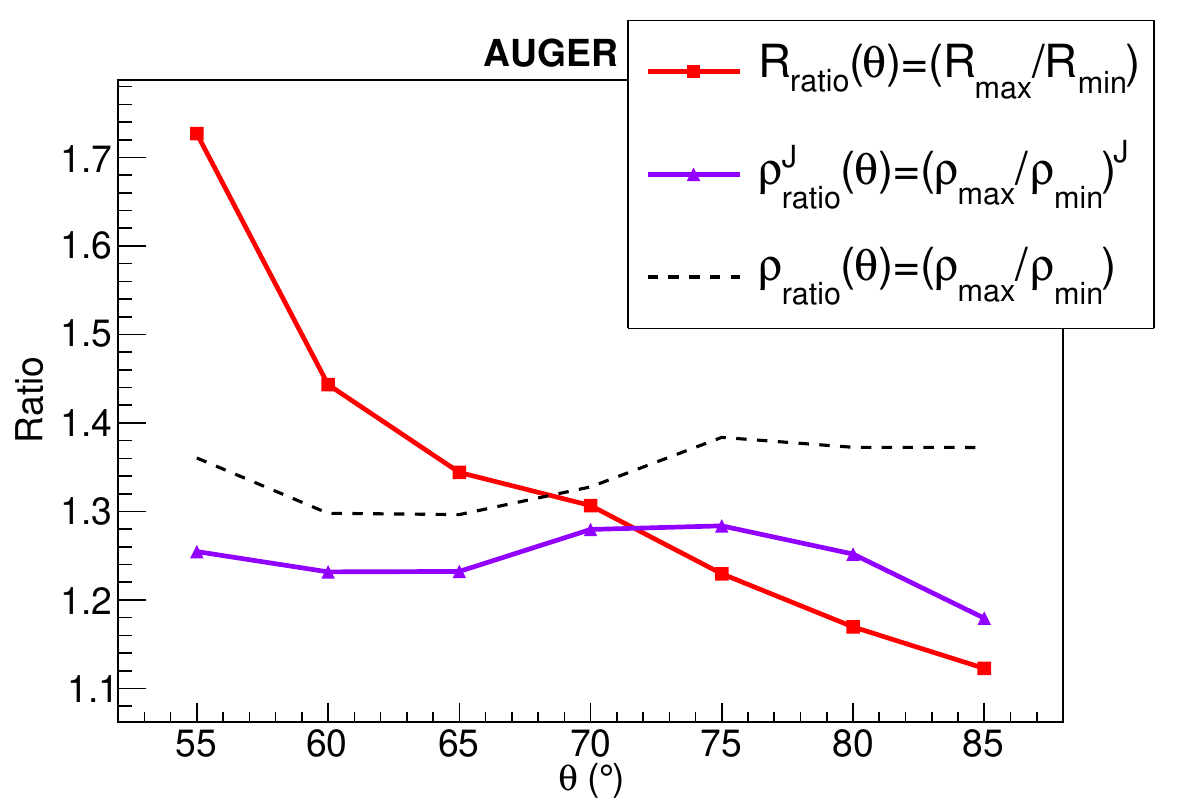}\includegraphics[width=0.5\textwidth]{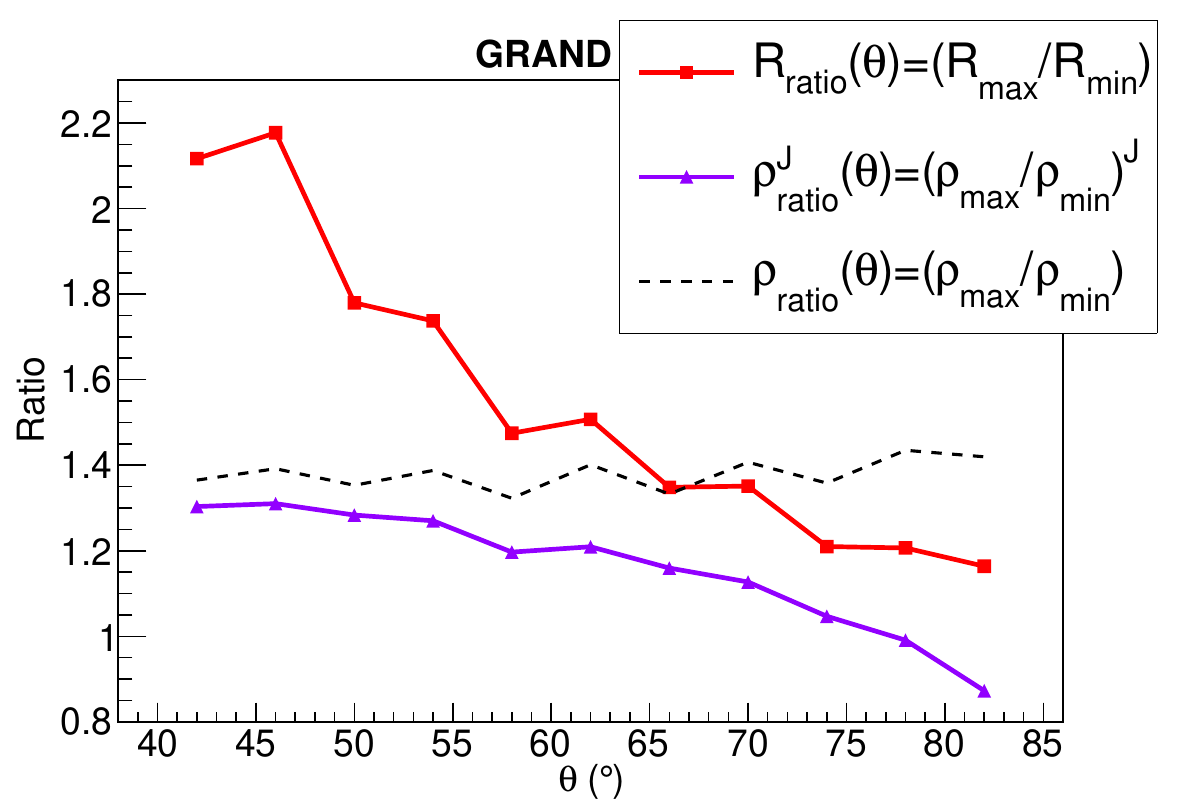}
  \includegraphics[width=0.5\textwidth]{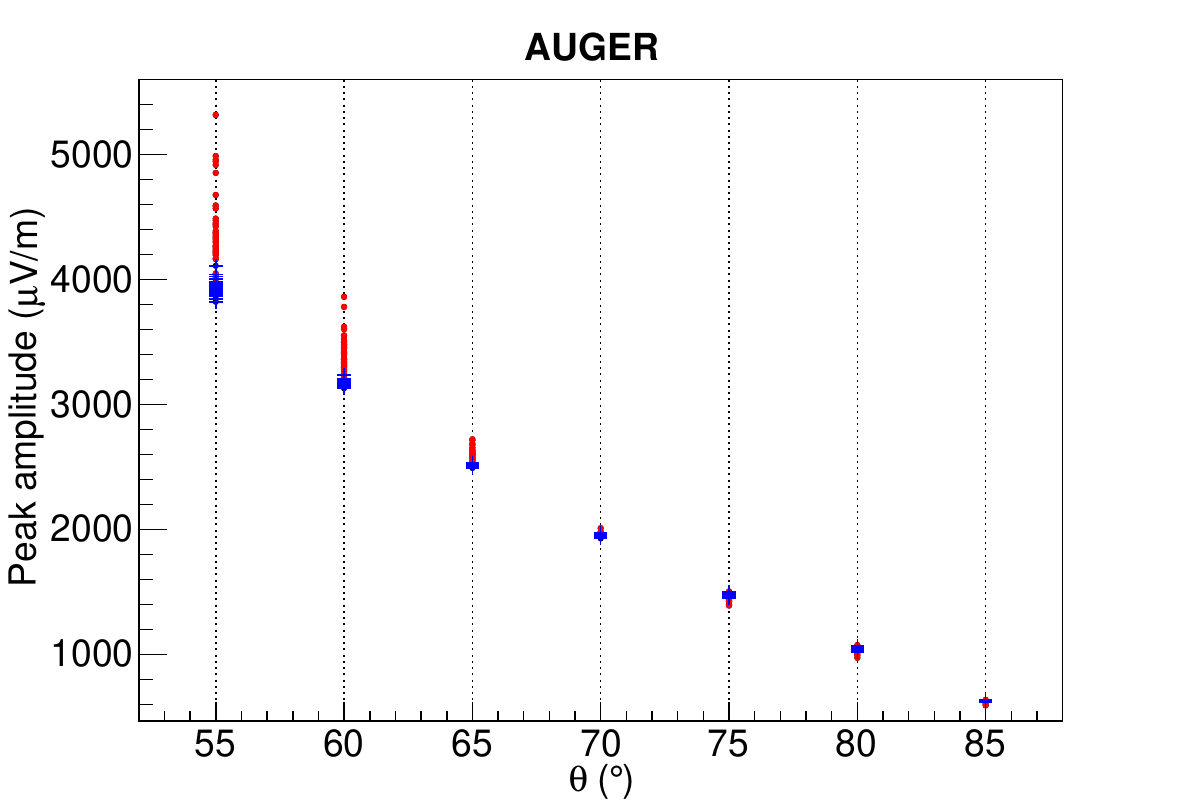}\includegraphics[width=0.5\textwidth]{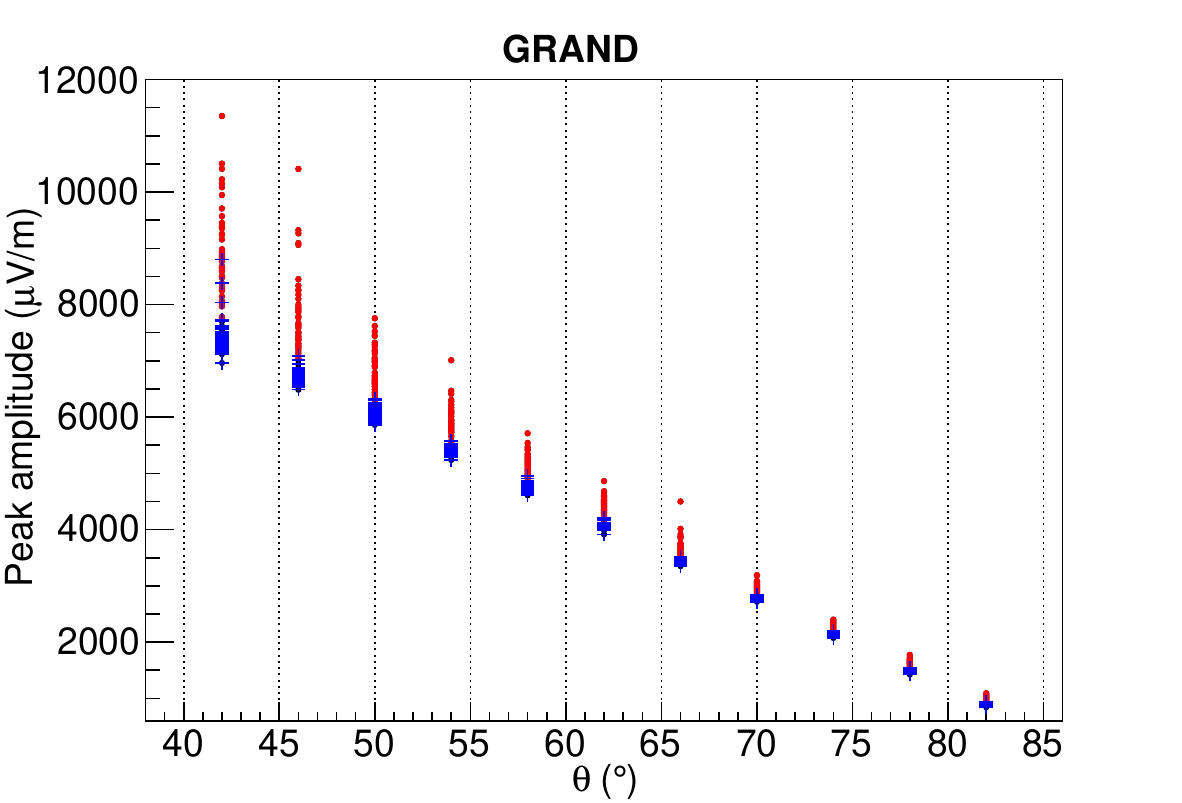}
  \includegraphics[width=0.5\textwidth]{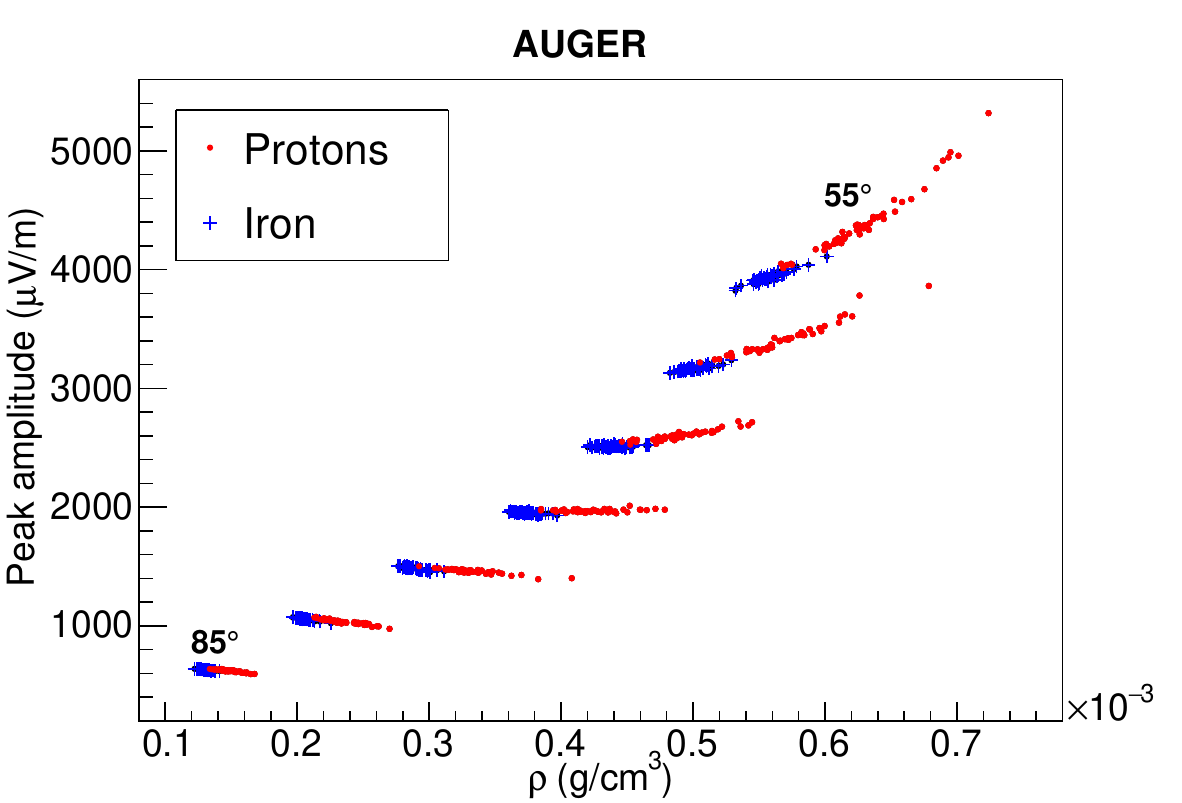}\includegraphics[width=0.5\textwidth]{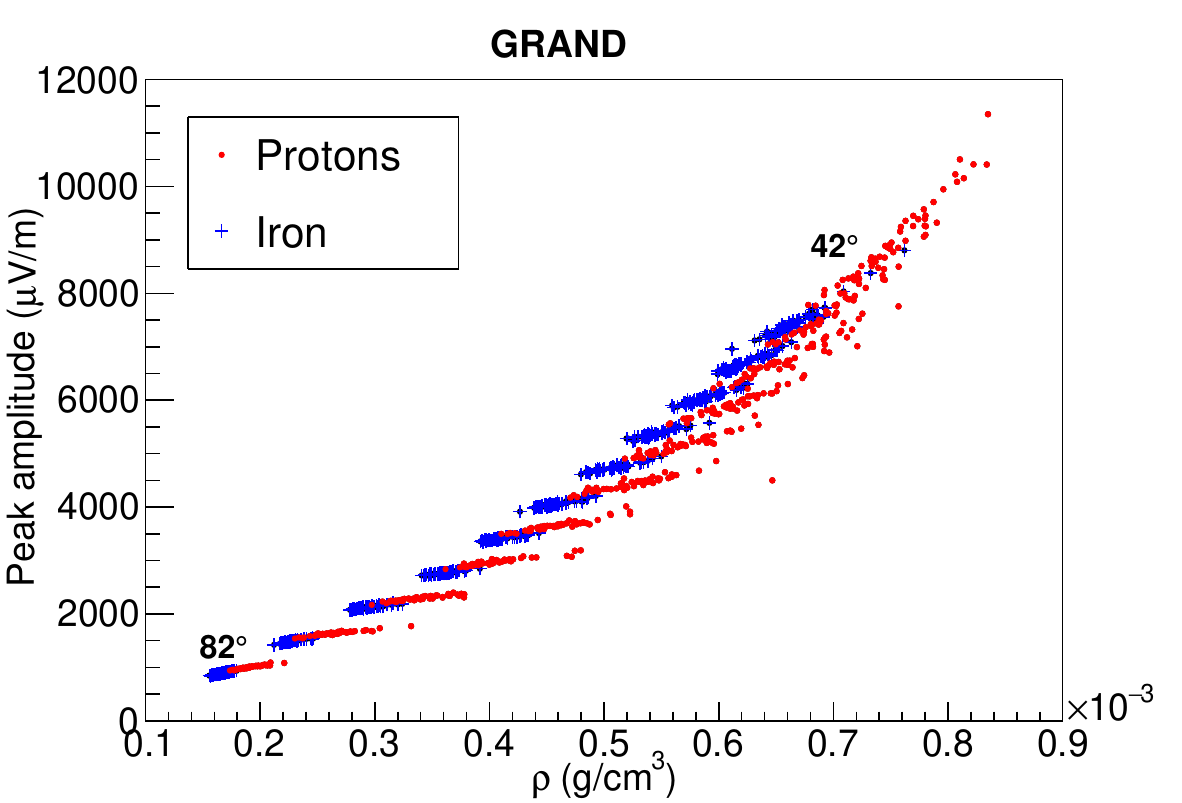}
  \caption{Top panels: Relative strength of the $1/R$ (red) and $(1/\rho)^{J(\theta)}$ (purple) scaling factors, as a function of zenith angle, for the Auger site (left) and  for the GRAND site (right). Also shown as a dashed line is the $(1/\rho)$ scaling, which does not take coherence loss ($J$) into account. Middle panels: Peak amplitude of the LDFs obtained from the full set of ZHAires simulations as a function of zenith angle for proton- (red dots) and iron-induced showers (blue crosses) for the Auger site (left) and GRAND site (right). Bottom panels: Same as the middle panels, but showing the peak amplitudes as a function of the air density at \xmax instead of the zenith angle. Top and bottom panels adapted from~\cite{RLDF-ICRC2025}. See text for more details.}
\label{fig:FactorRatios}
\end{center}
\end{figure}

We show these ratios as a function of zenith on the top panels of Fig.~\ref{fig:FactorRatios}. Red for the $1/R$ scaling and purple for the $(1/\rho)^{J(\theta)}$ scaling. Since these ratios represent the strength of each scaling, if the red line is above the purple one, the $1/R$ scaling dominates, increasing the field with \xmax. Conversely, if the red line is below, $(1/\rho)^{J(\theta)}$ dominates, decreasing the field with \xmax. At GRAND, the much larger geomagnetic field and loss of coherence make the $1/R$ scaling dominant at all zeniths, as can be seen on the top right panel of Fig.~\ref{fig:FactorRatios}. This means that at GRAND, regardless of zenith, proton-induced showers should tend to have, on average, higher electric fields than iron-induced ones. In contrast, at Auger (top left panel), the red and purple lines cross at around $72^\circ$, signaling a transition in the dominant scaling. Below this zenith, proton-induced showers should tend to have higher fields, while above it, iron-induced showers should dominate. One can also see that, in general (disregarding the $72^\circ$ Auger transition region), the difference between the two scalings tends to decrease with zenith angle at both sites. This suggests that as the zenith increases, we should expect smaller amplitude differences between proton- and iron-induced showers.

These predictions, solely based on the geometry of the shower and on our estimate for the $J(\theta)$ factor, match the results of the full simulations. On the middle and bottom panels of Fig.~\ref{fig:FactorRatios} we show the simulated maximum LDF amplitudes for iron- (blue crosses) and proton-induced showers (red dots), as a function of zenith angle (middle panels) and as a function of the density $\rho$ at \xmax (bottom panels). The middle panels highlight the behavior of the amplitudes at fixed zeniths, while in the bottom panels, the slope of the points for each zenith reflects the derivative of the amplitude with respect to $\rho$ (\xmax). On the middle right panel of Fig.~\ref{fig:FactorRatios} we show the simulation results for GRAND. One can see that proton showers tend to have higher electric fields at all zeniths, as predicted using the proposed scalings. In contrast, for Auger (middle left panel of Fig.~\ref{fig:FactorRatios}) we see that proton showers tend to have higher fields for zeniths below $72^\circ$, while iron showers dominate above that zenith. This matches the transition region predicted in the top left panel. Also, the size of the amplitude difference between proton- and iron-induced showers tends to diminish with increasing zenith for both sites, in agreement with the ratios shown on the top panels of Fig.~\ref{fig:FactorRatios}. This can be seen more clearly on the bottom panels. In the case of GRAND (bottom right panel), although protons tend to have higher fields at all zeniths, the inclination of the points for a given zenith decreases with increasing zenith. In the case of Auger (bottom left panel), this inclination also tends to decrease with zenith. But at $70^\circ$ (centered around $\rho=0.4\times10^{-3} g/cm^{3}$) the slope is almost horizontal, suggesting that the amplitude is almost independent of \xmax at that angle. Above $70^\circ$, the slope of the points reverses and iron-induced showers now tend to have the higher electric fields, matching the transition region.

\begin{figure}[!htb]
\begin{center}

\includegraphics[width=0.49\textwidth]{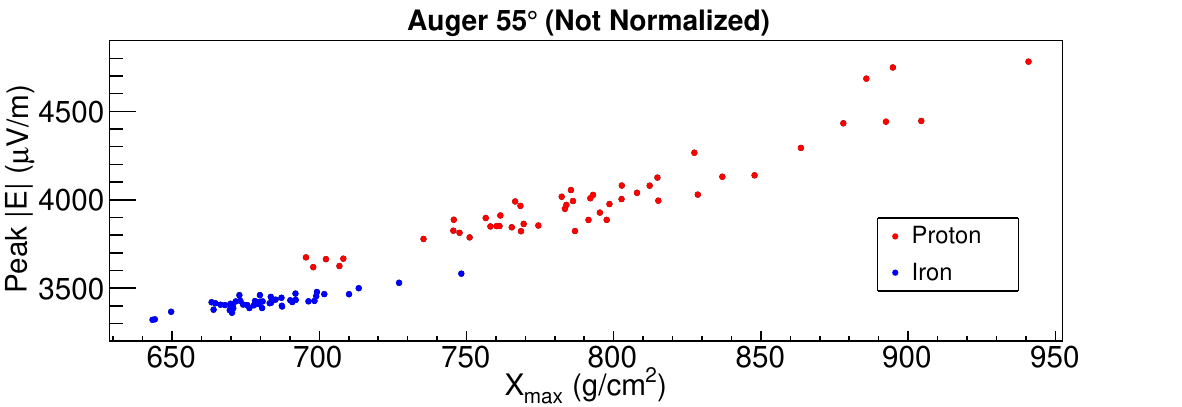}
\includegraphics[width=0.49\textwidth]{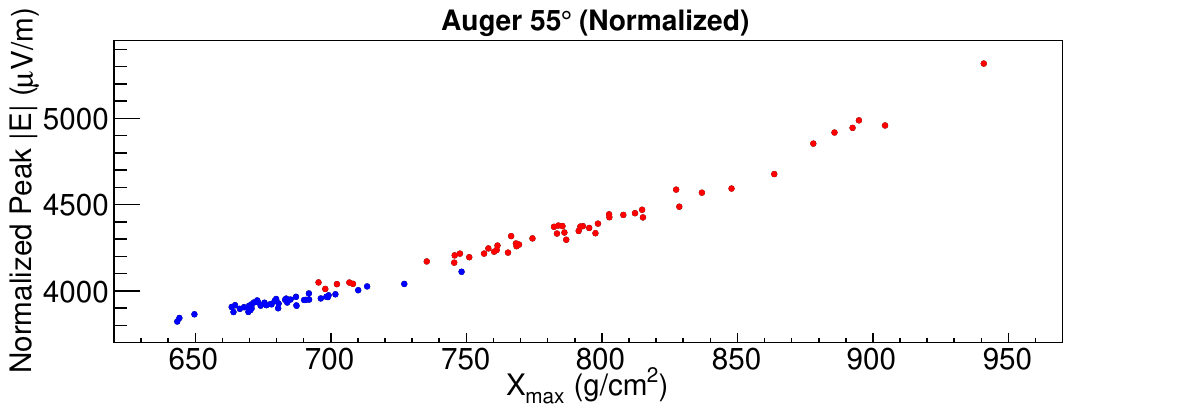}

\includegraphics[width=0.49\textwidth]{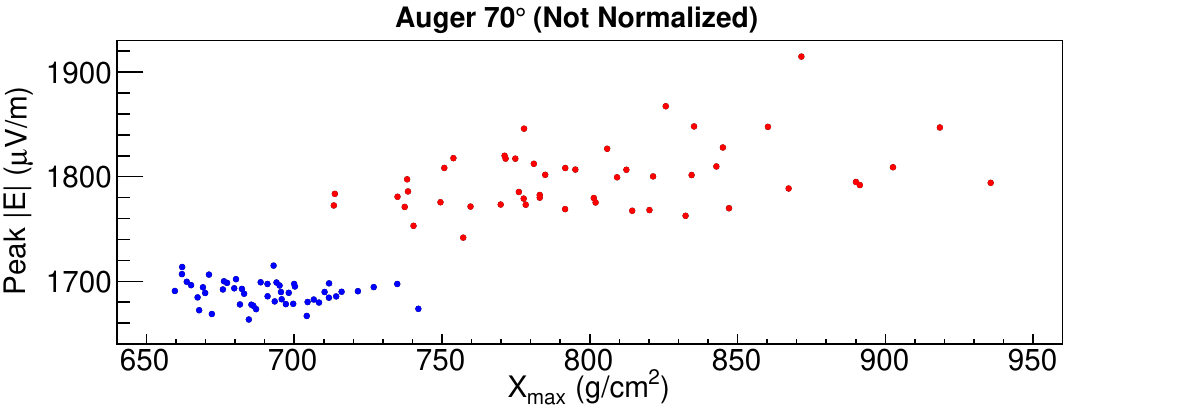}
\includegraphics[width=0.49\textwidth]{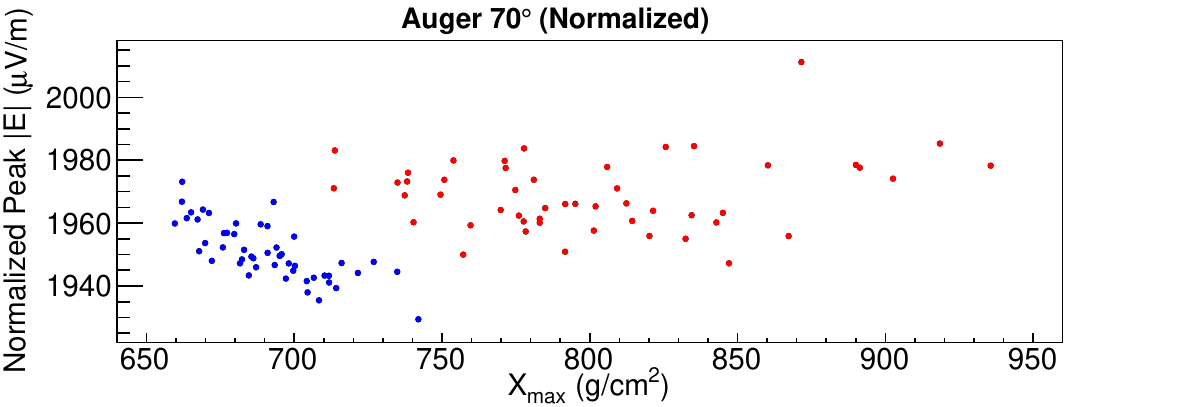}

\includegraphics[width=0.49\textwidth]{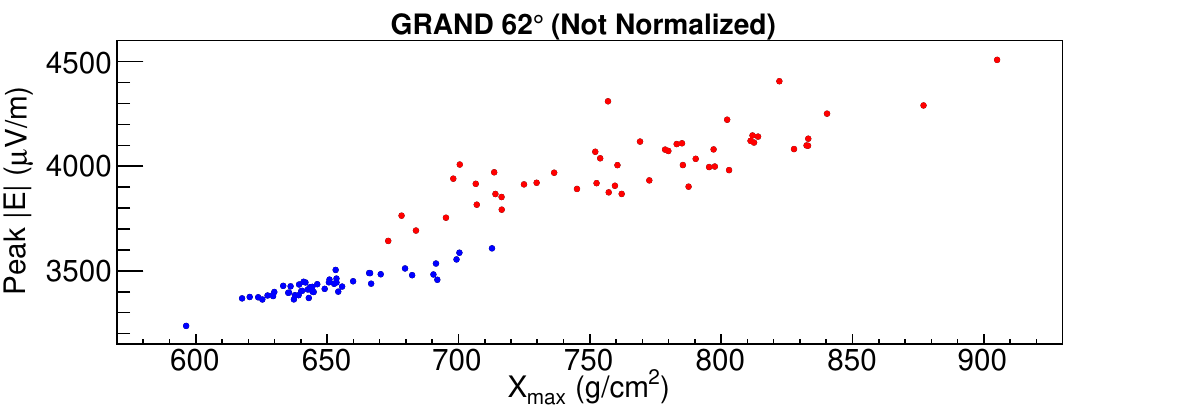}
\includegraphics[width=0.49\textwidth]{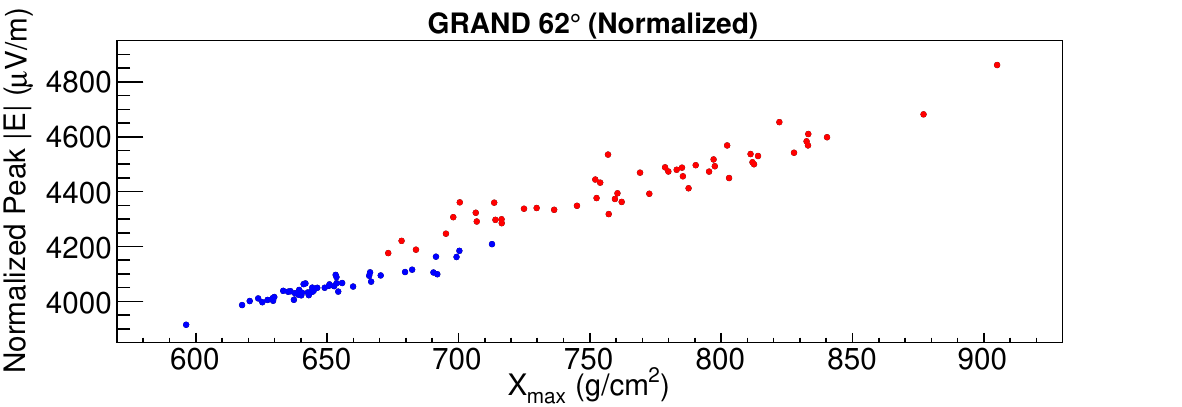}

\includegraphics[width=0.49\textwidth]{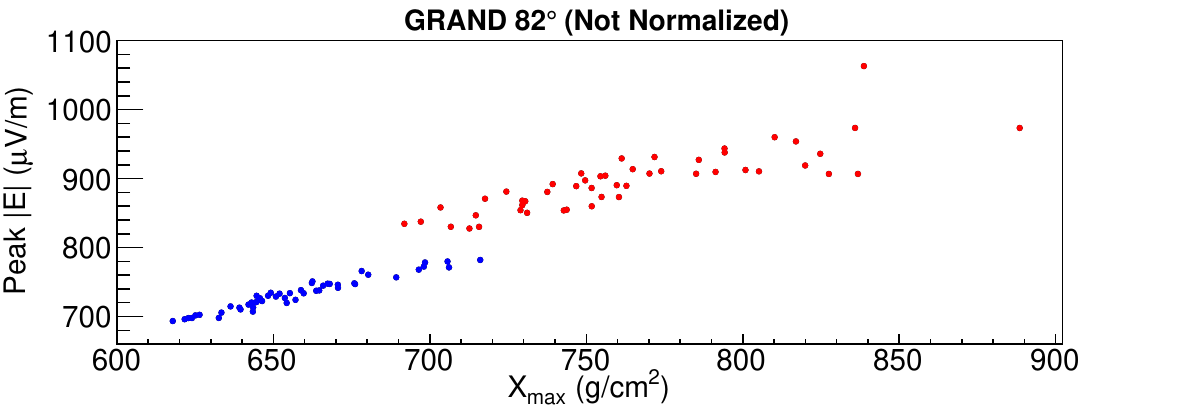}
\includegraphics[width=0.49\textwidth]{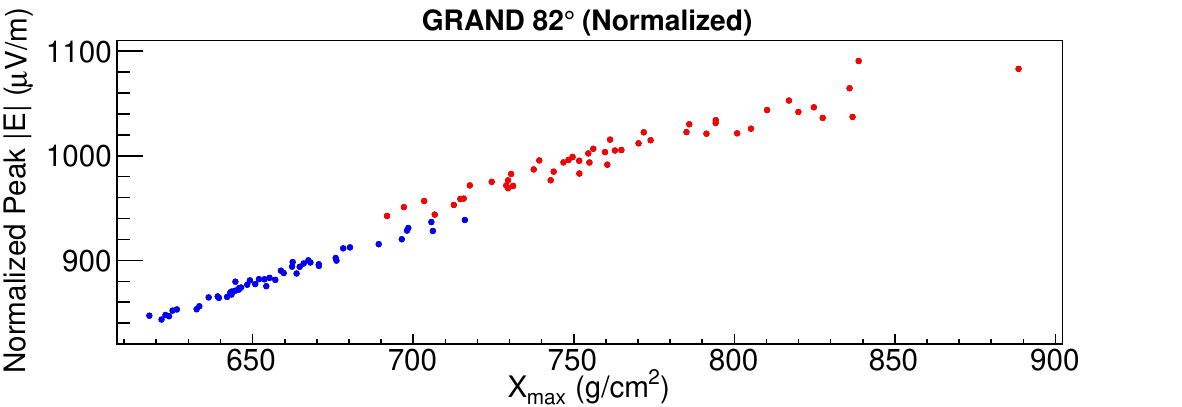}

\caption{Peak electric field amplitudes of the radio LDF (maximum LDF amplitudes) as a function of \xmax for proton- (red) and iron-induced (blue) showers at representative zenith angles at the Auger (top two rows) and GRAND (bottom two rows) sites. Left column: peak amplitudes without normalization. Right column: peak amplitudes normalized by the electromagnetic (EM) energy of each shower. Each panel is labeled by site and zenith angle. EM-energy normalization strongly reduces the composition dependence and reveals the dominant dependence of the maximum LDF amplitude on \xmax.}
\label{fig:SpeakVsXmax}

\end{center}
\end{figure}

This behavior can be seen more directly by looking at the maximum LDF amplitudes as a function of \xmax for fixed zenith angles. In Fig.~\ref{fig:SpeakVsXmax} we show the peak electric field amplitudes obtained from the simulations for proton- and iron-induced showers at several representative zenith angles at the Auger and GRAND sites, both before and after normalization by the EM energy of each shower. Before normalization, proton- and iron-induced showers show clear amplitude differences at fixed \xmax. After EM-energy normalization, these differences are strongly reduced and the peak amplitudes become primarily governed by \xmax, with only a small residual dependence on primary composition that varies with zenith angle. The reduction of the composition dependence is particularly evident at $\theta=55^\circ$ at Auger, while at $\theta=62^\circ$ at GRAND a somewhat larger residual remains. Near the Auger transition region ($\theta=70^\circ$), the peak amplitude becomes nearly independent of \xmax, as expected from the competing scalings discussed above. These results confirm that, after accounting for the EM energy, the dominant remaining dependence of the peak amplitude is on \xmax, with the strength of this dependence modulated by zenith angle and detector site. 

\section{Discussion and conclusion}
\label{sec:discussion}

Although the main goal of this work was just to understand what gives rise to the electric field amplitude dependence on \xmax, this dependence might have impacts on other radio reconstruction techniques. It is well known that the shape of the radio footprint at ground level depends on \xmax. This fact is used by \xmax reconstruction methods, such as the LOFAR-like ones~\cite{OriginalLofarXmax,LofarXmax2021,AERAXmaxPRL,AERAXmax}. But, as we have shown, \xmax also has an impact on the measured electric field amplitudes. This means that both, the shape and the brightness of the radio footprint have information on the \xmax of the shower.

The LOFAR-like \xmax reconstructions compare data with multiple simulations and use a $\chi^2$ method to obtain \xmax. These methods typically employ an energy scaling factor to calculate $\chi^2$, such as the factor $f_s$ of Eq. 1 of~\cite{LofarXmax2021} or the factor $S$ of Eq. 2 of~\cite{AERAXmax}. This factor scales the measured fluences to minimize the difference between the data and the multiple simulations. This is done to take into account energy measurement uncertainties and other systematic effects during the reconstruction. After the full analysis, one can obtain the value of the scaling factor that minimizes the differences between data and simulation, which in turn can provide a better energy estimate for the shower. Since the radio fluence is obtained from the electric field, we expect it to follow a similar dependence on \xmax as the peak amplitude discussed above.

However, leaving the scaling factor completely free for every shower, which is the case for the AERA \xmax reconstruction~\cite{AERAXmax}, also has a disadvantage. It will erase any amplitude (fluence) differences between the showers and thus lose some information on \xmax. In that case, the method will only pick up the footprint shape dependence on \xmax and not the amplitude dependence. This is especially critical for zeniths close to the “magic angle”, where the only \xmax dependence is on amplitude (see right panels of Fig.~\ref{fig:LDFExamples}). As a result, the method could yield very poor or even failed reconstructions in this zenith range. But, as far as we know, no one has applied the LOFAR method to very inclined showers yet. At lower zeniths, where the shape dependence on \xmax is stronger, the effect of a completely free scaling factor should be less pronounced. But constraining this scaling factor could in principle improve the \xmax resolution, since the method would then automatically take into account both, the shape and amplitude \xmax dependences.

In a similar vein, we previously developed an event-by-event composition discrimination method~\cite{compositionpaper} that uses a $\chi^2$ approach very similar to the LOFAR one. But instead of reconstructing \xmax, it tries to infer the composition directly. This method also features an energy scaling factor $f_s$ (see Eq. 1 of~\cite{compositionpaper}). If we leave $f_s$ free, any amplitude differences between the showers will be erased and some information on \xmax (composition) will be lost, just like in the LOFAR-like \xmax reconstructions discussed above. Although not understood at the time, this is the reason we observed a strong decrease in the composition discrimination accuracy when leaving the scaling factor $f_s$ free (compare Figs. 2 and 3 of~\cite{compositionpaper}), especially at higher zenith angles. As the zenith angle increases, the dependence of the footprint shape on \xmax diminishes, and the amplitude dependence on \xmax becomes the most important factor for composition discrimination, which will be completely erased if $f_s$ is set as a free parameter for each shower.

The strong dependence of the electric field amplitudes on \xmax discussed in this work may also have implications for the estimation of the EM energy of the shower. The total energy emitted in radio waves, which is commonly used to estimate the EM energy of the shower, can be obtained by integrating the squared electric field over time and over the footprint area. In the magic angle example shown in the right panels of Fig.~\ref{fig:LDFExamples}, the footprint area remains nearly unchanged while the electric field amplitudes vary with \xmax, suggesting that showers with the same EM energy could produce different total energies emitted in radio waves and therefore potentially lead to different estimates of the EM energy. However, the energy emitted in radio waves is usually determined from the radio fluence, which is the time integral of the squared electric field of the radio signal time trace (see Eq. 1 of~\cite{AERAXmax}). 
Since the pulse width is not expected to change significantly enough with \xmax to compensate for the amplitude variations, the resulting radio fluence could therefore, in principle, exhibit a similar dependence on \xmax. Nevertheless, this should be regarded as a qualitative statement, since the dependence of the radio fluence on \xmax has not been explicitly studied here. A dedicated study of this effect is deferred to future work. That said, it is interesting to note that some EM energy reconstruction methods based on the radio fluence, such as \cite{GlaserRadiationEnergy,FelixTimInclinedRec}, already include empirical corrections that account for the distance to and the air density at \xmax. The results presented here provide a physical interpretation of the origin of such corrections, showing that they arise naturally from the competing $(1/R)$ and $(1/\rho)^J$ scalings of the emission. Traditionally, such variations of the radio signal amplitude have mainly been treated as effects that must be corrected for in reconstruction methods. However, we believe that this amplitude dependence on \xmax should not only be regarded as a correction to be accounted for, but as a physical effect that carries information about the shower development, as illustrated in~\cite{MLpaper}.

As discussed above, if the radio fluence depends on \xmax, this raises the possibility that this dependence could introduce \xmax or composition biases in the EM energy reconstruction if not properly accounted for. A quantitative assessment of this effect would require both an explicit study of the dependence of the radio fluence on \xmax and a full detector level reconstruction analysis to evaluate its impact on the reconstructed EM energy, which is beyond the scope of this work.

Since the radio amplitude is also proportional to the EM shower energy, an estimate of this energy is required before the amplitude dependence on \xmax can be exploited for composition discrimination. Several methods exist~\cite{FelixTimInclinedRec,LukasInclinedICRC2025,AERAEMEnergy,GlaserRadiationEnergy} that reconstruct the EM energy from the radio signal alone, some with quoted uncertainties as low as $\sim5\%$~\cite{FelixTimInclinedRec}. Once the EM energy is estimated, the remaining amplitude variations primarily reflect differences in shower development and therefore carry information on the primary composition. A concrete implementation of this idea using ML techniques is presented in~\cite{MLpaper}.

We believe the peak amplitude \xmax dependence discussed in this work is a strong enough effect to allow for event-by-event composition discrimination. As we briefly discussed in section~\ref{sec:motivation}, the simple ML method that motivated this work exploited this amplitude dependence to discriminate between heavy and light primaries. Even at early stages of development it yielded very good accuracies, despite the very large energy uncertainties used. Since the start of this work, the ML method has been refined and expanded to also include spectral slopes for discrimination~\cite{MLpaper,RFDiscrimination-ICRC2025}. But even more refined methods that exploit this amplitude dependence on \xmax could be devised in the future, further increasing the accuracy of event-by-event primary discrimination.

While the results presented here reflect the strength of the \xmax amplitude dependence as obtained from our simulations, we emphasize that the actual strength of the effect in nature could be different. Also, the strengths obtained from different simulation codes and/or hadronic interaction models are expected to differ, as these are known to influence \xmax distributions and coherence effects~\cite{zhaires-air}. This is illustrated in Fig. 6 of~\cite{TimRadioRenaissance}, which compares results from two different simulations. Nevertheless, the physical explanation we provide, based on the competing $(1/\rho)^J$ and $(1/R)$ scalings, remains generally valid (see section~\ref{sec:PredictionsAndComparison}). It illustrates the underlying origin of the observed \xmax dependence and demonstrates that our work captures the essential physics of the effect, even if the precise magnitude may change under different simulation conditions.

An interesting, yet unexpected finding from our simulations is that the electric field amplitude of very inclined showers appears to be less sensitive to the geomagnetic field than at lower zeniths. This can be seen on the top left panel of Fig.~\ref{fig:Geo-Ask-emission}. Although the geomagnetic field at GRAND is much larger than at Auger, the amplitudes of the electric field at both sites are very similar above $\sim80^\circ$. This relates to the higher coherence loss induced by higher geomagnetic fields. This means that, for experiments focusing on radio detection of nearly horizontal events, the geomagnetic field at the site is less critical than for experiments at lower zeniths. So, for this type of experiment, a site with a low geomagnetic field seems to be a viable option. A more detailed study of the relation between the geomagnetic field and the electric field amplitudes for high zenith angle showers would be desirable. Such a study could establish limits for the lowest viable geomagnetic field as a function of zenith angle.

In this work we have shown that there is a strong dependence of the radio LDF electric field amplitudes on the position of \xmax in the atmosphere, even when accounting for differences in the EM energy of the showers. This \xmax dependence, which also leads to a primary composition dependence, was explained in terms of two competing scalings of the measured electric field: One goes with $(1/\rho)^J$, where $\rho$ is the air density at \xmax and $J$ is a zenith dependent non-linearity factor describing coherence loss. This density scaling tends to decrease the geomagnetic emission of deeper showers. The other scaling goes with $(1/R)$, where $R$ is the distance from \xmax to the core at ground, and instead increases the measured electric field of deeper showers. At low zenith angles, solely due to shower geometry, the $(1/R)$ scaling is stronger and leads to larger measured electric fields in the case of the deeper showers induced by lighter primary compositions. The picture at higher zenith angles, i.e., lower densities, is more nuanced. In this region, the deflections due to the Lorentz force are much larger and increase the perpendicular momenta of $e^\pm$, which would tend to increase the radio emission. However, it also increases the time delays between the particle tracks, decreasing the coherence of the emission. This loss of coherence is highly dependent on the strength of the Lorentz force and therefore on the geomagnetic field. It can slow down, or even reverse, the increase of the radio emission with decreasing air density. This effect is represented by the non-linearity factor $J(\theta)$, which determines whether lighter primary compositions will induce higher or lower fields than heavy primaries at these higher zenith angles. This strong LDF amplitude dependence on \xmax/composition can be used to directly infer, even bypassing any \xmax reconstruction, the cosmic ray primary composition on an event-by-event basis. It could also be used to improve \xmax reconstructions in LOFAR-like methods, particularly at high zenith angles. This could be accomplished by keeping the energy scaling factor, normally used when comparing data with simulations, fixed. This amplitude dependence, if not taken into account, raises the possibility that EM energy estimates could exhibit an \xmax/composition bias. 

\section{Acknowledgments}
We would like to thank Jaime Alvarez-Muñiz and Bjarni Pont for their many valuable suggestions and for proofreading the manuscript.
We acknowledge the Polish National Agency for Academic Exchange within Polish Returns Program no. PPN/PPO/2020/1/00024/U/00001 and the National Science Centre Poland for NCN OPUS grant no. 2022/45/B/ST2/02889. Disclosure: During the final editing of the manuscript, we used the OpenAI language model ChatGPT (GPT-5) as a tool to assist with checking grammar, typos, clarity and bibliographic formatting. All text was written and revised by the authors, who take full responsibility for the content of the article.


\end{document}